\documentclass[final,3p]{elsarticle}
\makeatletter 
\def\ps@pprintTitle{%
 \def\@oddfoot{\footnotesize\itshape
       To be published in \ifx\@journal\@empty Elsevier
       \else\@journal\fi\hfill\today}}
\makeatother

\newif\ifBeamerClass
\BeamerClassfalse
\newif\ifEnableListings
\EnableListingsfalse
\usepackage{amsmath,amsfonts,amssymb} % for maths
\usepackage{mathtools} % contains amsthm
\usepackage{upgreek} % to get greek letters in upright mode, like \upphi
\usepackage{easybmat} % to easily create large block matrices
\usepackage{fixmath} % provides upercase greek: \upxi,\upeta,...
\usepackage[font=footnotesize,list=true]{subcaption} % Needed for subfigures
\usepackage{booktabs} % needed for \toprule, \bottomrule,... in tables
\usepackage[exponent-product=\cdot]{siunitx} % for nice SI unit formatting

\usepackage{bm} % to get nice symbols in boldface mode
\usepackage{xcolor}
\usepackage{microtype}
\usepackage{lmodern}

\newcommand{\graphicsFolder}{graphics}

\renewcommand{\vec}[1]{\mathbold #1} % ISO80000-2 format
\newcommand{\besselj}{\mathrm{j}} 	% ISO80000-2 format
\newcommand{\bessely}{\mathrm{y}} 	% ISO80000-2 format
\newcommand{\besselJ}{\mathrm{J}} 	% ISO80000-2 format
\newcommand{\besselY}{\mathrm{Y}}  	% ISO80000-2 format
\newcommand{\Hankel}{\mathrm{H}}   	% ISO80000-2 format
\newcommand{\legendre}{\mathrm{P}} 	% ISO80000-2 format
\newcommand{\hankel}{\mathrm{h}}   	% ISO80000-2 format
\newcommand{\heaviside}{\mathrm{H}} % ISO80000-2 format
\newcommand{\diff}{\mathrm{d}} % For differentials
\newcommand{\idiff}{\, \mathrm{d}} % For differentials in integrals.
\newcommand{\zerovec}{\mathbf{0}}
\newcommand{\transpose}{\top}

\newcommand{\R}{\mathbb{R}}
\providecommand{\C}{\mathbb{C}}

\newcommand{\N}{\mathbb{N}}
\newcommand{\PI}{\uppi}
\newcommand{\euler}{\mathrm{e}}
\newcommand{\imag}{\mathrm{i}}

\providecommand*{\pderiv}[3][]{\frac{\partial^{#1}#2}{\partial #3^{#1}}}
\providecommand*{\deriv}[3][]{\frac{\diff^{#1}#2}{\diff #3^{#1}}}

\renewcommand{\leq}{\leqslant}

\DeclareMathSymbol{\GAMMA}{\mathalpha}{operators}{0}

\DeclareMathOperator{\fourier}{\mathcal{F}}
\DeclareMathOperator{\TS}{TS}

\DeclareMathOperator{\Ai}{Ai}
\DeclareMathOperator{\Bi}{Bi}

\renewcommand\Re{\operatorname{Re}}
\renewcommand\Im{\operatorname{Im}}

\newcommand\realmin{\operatorname{REALMIN}}
\newcommand\realmax{\operatorname{REALMAX}}

\renewcommand{\Xi}{{\vec{t}_1}}

\newcommand{\energyNorm}[2]{%
  {\left\vert\kern-0.25ex\left\vert\kern-0.25ex\left\vert #1 
    \right\vert\kern-0.25ex\right\vert\kern-0.25ex\right\vert}_{#2}
}

\let\originalleft\left
\let\originalright\right
\renewcommand{\left}{\mathopen{}\mathclose\bgroup\originalleft}
\renewcommand{\right}{\aftergroup\egroup\originalright}

\newcommand{\st}{\,:\,}

\usepackage{xspace}
\newcommand{\MATLAB}{\textsc{Matlab}\xspace}

\usepackage[hidelinks,pdftex]{hyperref} % needed for links
\hypersetup{colorlinks = true, citecolor=blue, linkcolor=blue, urlcolor=blue} % Color the links
\usepackage[noabbrev]{cleveref} % To have easy referencing for figures, equations, ... noabbrev option can be used for full name
\usepackage{longtable}
\usepackage{enumitem}

\newtheorem{remark}{Remark}
\graphicspath{{./graphics/}}
\journal{Journal of Sound and Vibration}

\begin{document}
\begin{frontmatter}

\title{Generalization and stabilization of exact scattering solutions\\for spherical symmetric scatterers}

\author[venas]{\texorpdfstring{Jon Vegard Ven{\aa}s\corref{cor}}{Jon Vegard Ven{\aa}s}}
\ead{JonVegard.Venas@sintef.no}

\author[jenserud]{Trond Jenserud}
\ead{Trond.Jenserud@ffi.no}

\address[venas]{SINTEF Digital, Mathematics and Cybernetics, 7037 Trondheim, Norway}
\address[jenserud]{Norwegian Defence Research Establishment (FFI), Horten NO-3191, Norway}
\cortext[cor]{Corresponding author.}

\begin{abstract}
In a previous work the authors described a fast high-fidelity computer model for acoustic scattering from multi-layered elastic spheres. This work is now extended with a scaling strategy significantly mitigating the problem of overflow and thus expanding the useful frequency range of the model. Moreover, new boundary conditions and loads are implemented. Most important are the fluid-fluid and solid-solid couplings, which allow a completely general layering of the scattering object. Sound hard and sound soft boundary conditions are implemented for solids and fluids respectively. In addition to the existing acoustic excitation, mechanical excitation in the form of point-excitation and surface excitation are implemented.  Attenuation in the form of hysteresis damping as well as viscous fluid layers are also included. Several numerical examples are included, with the purpose of validating the code against existing reference solutions. The examples include air bubbles, a coated steel shell and a point-exited steel shell.
\end{abstract}


\end{frontmatter}
\section{Introduction}
Computer models of acoustic scattering from three-dimensional elastic objects including internal structure are important tools in underwater acoustics, as well as in many other fields such as non-destructive testing, seismology, noise control and medical imaging.

In underwater acoustics such scattering models allow the study of realistic scattering scenarios, both from marine life~\cite{Stanton1998dbs,Stanton2000asb} and man-made objects~\cite{Sessarego1998sba,Williams2010asf,Lim2000asb,Plotnick2015hfb,Tesei2002mam}. Present work studies scattering by objects in the presence of interfaces, which enable modeling of partially and completely buried objects.
Being able to predict the acoustic response of objects aids in the development of methods for detection and identification, and reduces the need for expensive experiments~\cite{Werby2002ass,Bucaro2008bas,Espana2014asf,Jia2017rae,Sternini2020bso}.
Another application is noise and vibration reduction, where sandwich structures with sound absorbent cores are used to improve the acoustic insulation~\cite{Seilsepour2022aic,Zarastvand2021iot,Zarastvand2022awt}.

Many computational techniques have been developed over the years for studying scattering from fluid-loaded objects, ranging from fast approximate methods to computer intensive high-fidelity models. In the development of such methods, reference models are an invaluable tool for validating the accuracy and valid ranges of the different methods~\cite{Venas2018iao,Venas2020ibe}.

Exact solutions are only known for a few simple shapes: the infinitely long cylinder and the sphere. There is therefore a need for reference models for more complex objects of simple shapes included inside layered acoustic media; for this purpose, we have developed an open sourced\footnote{The \textit{e3Dss} toolbox is available at \href{https://github.com/Zetison/e3Dss}{GitHub}.} toolbox for scattering by multi-layered elastic spheres~\cite{Venas2019e3s}. The model computes scattering by an elastic object submerged in an infinite fluid medium, waveguide effects are not considered. The model is exact in the sense that it is based on analytical expressions but requires numerical calculations to compute a truncated series with accuracy depending only on the precision used.

\begin{figure}
	\centering
	\includegraphics{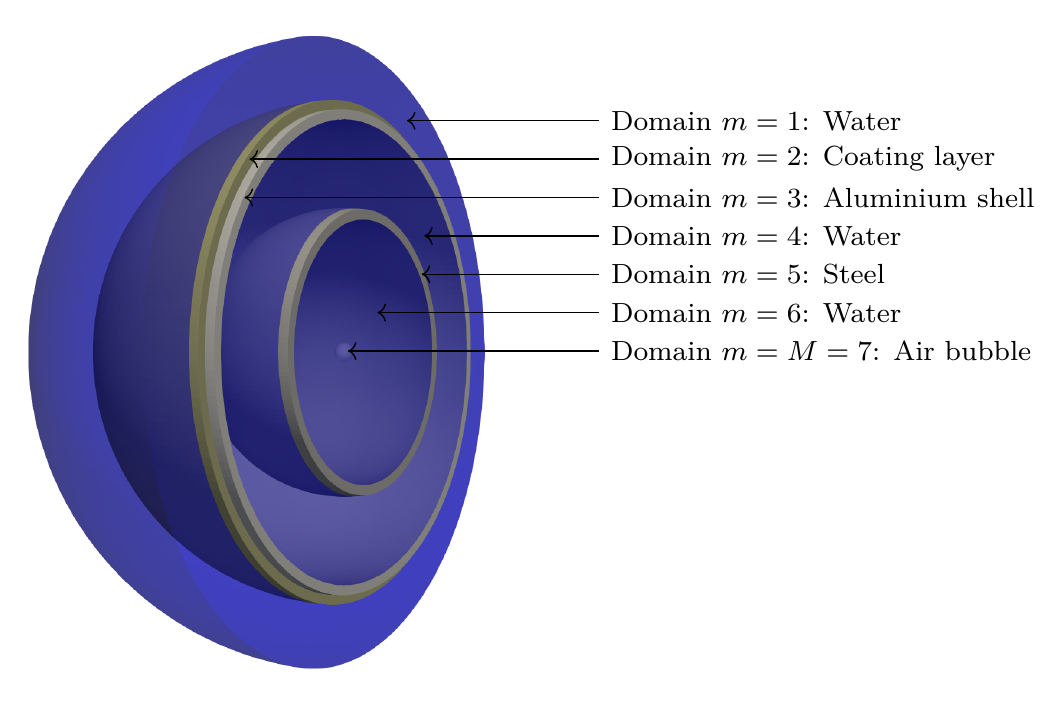}
	\caption{An outer spherical aluminum shell with a coating layer is immersed in water (unbounded domain), the interior being flooded with water and containing another steel shell also being flooded with water but has also an air bubble in its center. This illustrates some of the generalization being made; both fluid-fluid and solid-solid couplings are made for the multi layered model.}
	\label{Fig1:illustration}
\end{figure}

The present work solves some limitations of the model described in~\cite{Venas2019e3s} and extends its capabilities. New boundary conditions are implemented, including fluid-fluid and solid-solid coupling (e.g. as in~\Cref{Fig1:illustration}); this removes the former restriction on the layering to be alternating fluid and solid, such that completely general layering is now allowed. 
Sound hard and sound soft boundary conditions are also implemented, allowing perfectly rigid and pressure release domains to be modelled.
In order to simulate realistic materials attenuation needs to be included, and attenuation in the form of viscous fluid layers and hysteresis damping are implemented for this purpose. 
Source types now include both acoustical (plane wave and point source) and mechanical (point force and surface) excitation.

Another issue that is addressed is the overflow problem that occurs at high $ka$-numbers due to the exponential behaviour of Bessel functions for large order. The resulting instabilities may hide significant physics and render the solution useless. A scaling strategy is applied that mitigates this problem and expands the useful frequency range significantly.

The novelty of this work is thus the stabilization and generalization of the work in~\cite{Venas2019e3s} both in terms of domain composition but also type of medium and boundary conditions. Besides some small alternation to the notations and the inclusion of scaling functions (for stabilization) the new scheme is similar to that of~\cite{Venas2019e3s}. The main achievement in this work is arguably the stabilization as this scaling is completely absent in the literature. The main motivation for this work (i.e. new boundary conditions) is to enable verification of numerical codes although the physical problem in and of itself is of interest.

The article is organized as follows:
\Cref{Eq1:generalSolution} introduces full Navier-Navier coupling conditions allowing completely general layering and implements viscous fluid layers through the linearized Navier-Stokes equation. 
The problem of overflow is treated in~\Cref{Sec1:besselLargOrder}, where a scaling strategy is implemented to mitigate the exponential behavior. The resulting stable solutions are given in~\Cref{Sec:stableSol}. 
Several numerical examples are given in~\Cref{Sec1:resultsDisc}: Scattering by gas bubbles in a liquid; empty spherical steel shell with soft coating; spherical steel shell with point excitation; and viscoelastic sphere. The examples serve to validate the implemented code by a comparison with existing benchmark solutions. 
\section{General solution}
\label{Eq1:generalSolution}
The restriction in~\cite{Venas2019e3s}, that the scattering object needs to consist of alternating fluid and solid domains is removed.
This forces us to alter the notation slightly. Denote by $M$ the number of media (solids and fluids) present such that we have $M-1$ spherical \textit{interfaces} at radii $R_m$, $m=1,\dots,M$, separating the solid and fluid layers (where $R_m > R_{m+1}$). If the innermost media has support at the origin, $R_M=0$. The domain number $m$ is denoted by
\begin{equation}
    \Omega_m = \{\vec{x}\in\R^3\st R_m\leq|\vec{x}| < R_{m-1}\}
\end{equation}
with the convention $R_0\to\infty$ for unbounded domains. We will also generalize the fluid domain to be viscous using linearized equations of continuity and the Navier-Stokes equation for a barotropic, viscous non-heat-conducting compressible fluid~\cite{Hasheminejad1993mif,Hasheminejad2005asf,Skelton1997tao}
\begin{align}
    \pderiv{\breve{\rho}'_m}{t} + \rho_m\nabla\cdot\breve{\vec{v}}_m &=0\label{Eq:continuity}\\
    \rho_m\pderiv{\breve{\vec{v}}_m}{t}+\nabla \breve{p}_m &= \mu_m\nabla^2\breve{\vec{v}}_m + \left(\frac13\mu_m + \mu_{\mathrm{b},m}\right)\nabla(\nabla\cdot\breve{\vec{v}}_m)\label{Eq:NavierStokes}\\
    \breve{p}_m &= c_m^2\breve{\rho}_m'\label{Eq:Barotropic}
\end{align}
where $\breve{p}_m$ represents the deviation of pressure from its mean value, $\breve{\vec{v}}_m$ is the velocity, $\breve{\rho}'_m$ is the time-varying part (as opposed to the constant equilibrium density $\rho_m$) of the total mass density, $\breve{\rho}_{\mathrm{tot},m}=\rho_m+\breve{\rho}'_m$, and $\mu_m$ and $\mu_{\mathrm{b},m}$ are the shear and the bulk coefficient of viscosity, respectively. Finally, $c_m$ is the ideal speed of sound evaluated at ambient conditions. \Cref{Eq:continuity,Eq:NavierStokes,Eq:Barotropic} can be combined into a single equation for the velocity field
\begin{equation}\label{Eq:fluidVelocityEq}
    \rho_m\pderiv[2]{\breve{\vec{v}}_m}{t} - \rho_m c_m^2\nabla(\nabla\cdot\breve{\vec{v}}_m) = \mu_m\nabla^2\pderiv{\breve{\vec{v}}_m}{t}+\left(\frac13\mu_m + \mu_{\mathrm{b},m}\right)\nabla\left(\nabla\cdot\pderiv{\breve{\vec{v}}_m}{t}\right).
\end{equation}
The time-dependent velocity field, $\breve{\vec{v}}_m$, can be written in terms of the time-dependent displacement field, $\breve{\vec{u}}_m$, by
\begin{equation}\label{Eq:u_vs_v}
    \breve{\vec{v}}_m=\pderiv{\breve{\vec{u}}_m}{t}.
\end{equation}
The displacement field, $\breve{\vec{u}}_m$, in the solid domain is governed by (as in~\cite{Venas2019e3s})
\begin{equation}\label{Eq1:navierTime}
	\rho_m\pderiv[2]{\breve{\vec{u}}_m}{t} = G_m\nabla^2\breve{\vec{u}}_m+\left(K_m+\frac{G_m}{3}\right)\nabla(\nabla \cdot\breve{\vec{u}}_m),
\end{equation}
where the \textit{bulk modulus}, $K_m$, and the \textit{shear modulus}, $G_m$, are defined by the Young's modulus, $E_m$, and Poisson's ratio, $\nu_m$, as
\begin{equation}
	K_m = \frac{E_m}{3(1-2\nu_m)}\quad\text{and}\quad G_m = \frac{E_m}{2(1+\nu_m)}.
\end{equation}
\Cref{Eq:fluidVelocityEq,Eq1:navierTime} in the frequency domain are obtained through Fourier transform\footnote{For a time dependent function $\breve{\Psi}(\vec{x},t)$, its Fourier transform is~\cite{Venas2019e3s}
\begin{align*}
	\Psi(\vec{x},\omega) = \left(\fourier\breve{\Psi}(\vec{x},\cdot)\right)(\omega) &= \int_{-\infty}^\infty \breve{\Psi}(\vec{x},t)\euler^{\imag \omega t}\idiff t
\end{align*}
where $\omega=2\PI f$ is the angular frequency, $f$ the frequency and $\imag=\sqrt{-1}$ is the imaginary unit. That is, we use the \textit{breve} notation, $\breve{\cdot}$, for the time-dependent Fourier transforms of functions in the frequency domain.} as
\begin{equation}\label{Eq1:navierFluid}
	\mu_m\nabla^2\vec{v}_m+\left(\frac13\mu_m + \mu_{\mathrm{b},m}+\frac{\imag \rho_mc_m^2}{\omega}\right)\nabla(\nabla \cdot\vec{v}_m) + \imag\omega\rho_m\vec{v}_m = \zerovec.
\end{equation}
and
\begin{equation}\label{Eq1:navierSolid}
	G_m\nabla^2\vec{u}_m+\left(K_m+\frac{G_m}{3}\right)\nabla(\nabla \cdot\vec{u}_m) +\rho_m\omega^2\vec{u}_m = \zerovec.
\end{equation}
respectively. \Cref{Eq1:navierFluid} can be written in terms of the fluid displacement using the Fourier transformation of \Cref{Eq:u_vs_v} 
\begin{equation}\label{Eq:u_vs_v_freq}
    \vec{v}_m = -\imag\omega\vec{u}_m
\end{equation}
such that
\begin{equation}\label{Eq1:navierFluidFreq}
	-\imag\omega\mu_m\nabla^2\vec{u}_m-\imag\omega\left(\frac13\mu_m + \mu_{\mathrm{b},m}+\frac{\imag \rho_m c_m^2}{\omega}\right)\nabla(\nabla \cdot\vec{u}_m) + \rho_m\omega^2\vec{u}_m = \zerovec.
\end{equation}
For fluid domains we define
\begin{equation}
   K_m \leftarrow \rho_m c_m^2 - \imag \omega\mu_{\mathrm{b},m}\quad\text{and}\quad  G_m \leftarrow -\imag\omega\mu_m
\end{equation}
which makes~\Cref{Eq1:navierFluidFreq} equivalent to~\Cref{Eq1:navierSolid}. This enables us to use same expressions for fluid and solid domain when expressing the general solution.

The solution for the displacement field in domain $m$, may be decomposed as\footnote{We use displacement-based potentials also for the fluid as was done in~\cite{Jenserud1990ars} (in contrast with the velocity based potentials used in~\cite{Skelton1997tao,Hasheminejad2005asf}).}
\begin{equation}\label{Eq1:LameSolution}
	\vec{u}_m = \nabla\phi_m + \nabla\times\vec{\psi}_m,\quad \vec{\psi}_m=\psi_{\upvarphi,m}\vec{e}_\upvarphi,
\end{equation}
where the displacement potentials $\phi_m$ and $\vec{\psi}_m$ solves
\begin{align}
	&\nabla^2\phi_m + a_m^2\phi_m = 0\label{Eq1:phiHelmholtz}\\
	&\nabla^2\vec{\psi}_m + b_m^2\vec{\psi}_m = \zerovec,\label{Eq1:PsiHelmholtz}
\end{align}
respectively. The angular wave numbers $a_m$ and $b_m$ are given by
\begin{align}
    a_m &= \frac{\omega}{c_{\mathrm{s},1,m}},\quad c_{\mathrm{s},1,m} = \sqrt{\frac{3K_m+4G_m}{3\rho_m}}\\
    b_m &= \frac{\omega}{c_{\mathrm{s},2,m}},\quad c_{\mathrm{s},2,m} = \sqrt{\frac{G_m}{\rho_m}}.
\end{align}
For non-viscous fluids we use the classical notation for the wave number, $k_m = \frac{\omega}{c_m} = a_m$, as $c_m=c_{\mathrm{s},1,m}$ in this case.
\begin{remark}
The angular wave numbers may for fluid domain be written in terms of the fluid parameters as\footnote{Note that~\cite{Hasheminejad1993mif,Hasheminejad2005asf} use the (linear Taylor expansion) approximation $a_m\approx \frac{\omega}{c_m}\left(1+\imag\frac{\omega\mu_m}{2\rho_m c_m^2}\left(\frac43 +\frac{\mu_{\mathrm{b},m}}{\mu_m}\right)\right)$.}
\begin{align}
    a_m &= \frac{\omega}{c_m}\left[1-\frac{\imag\omega\mu_m}{\rho_m c_m^2}\left(\frac{4}{3} + \frac{\mu_{\mathrm{b},m}}{\mu_m}\right)\right]^{-\frac{1}{2}}\\
    b_m &= \sqrt{\frac{\imag\omega\rho_m}{\mu_m}}.
\end{align}
\end{remark}
The following relations are obtained from~\Cref{Eq1:LameSolution,Eq:u_vs_v_freq,Eq1:phiHelmholtz}
\begin{align}
    \nabla\cdot\vec{u}_m &= -a_m^2\phi_m\\
    \nabla\cdot\vec{v}_m &= \imag\omega a_m^2\phi_m
\end{align}
and using~\Cref{Eq:continuity} we have for fluid domains
\begin{equation}
    p_m = \rho_m\omega^2\phi_m.
\end{equation}
The displacement is given in spherical coordinates as $\vec{u}_m=u_{\mathrm{r},m}\vec{e}_{\mathrm{r}}+u_{\upvartheta,m}\vec{e}_\upvartheta$ with components
\begin{equation}
	u_{\mathrm{r},m} = \pderiv{\phi_m}{r} + \frac{1}{r}\pderiv{\psi_{\upvarphi,m}}{\vartheta} + \frac{1}{r}\psi_{\upvarphi,m}\cot\vartheta,\qquad u_{\upvartheta,m} = \frac{1}{r}\pderiv{\phi_m}{\vartheta} -\pderiv{\psi_{\upvarphi,m}}{r} -\frac{1}{r}\psi_{\upvarphi,m}.
\end{equation}
As in~\cite{Venas2019e3s} \Cref{Eq1:phiHelmholtz,Eq1:PsiHelmholtz} can be written in spherical coordinates as
\begin{align}
	\pderiv{}{r}\left(r^2\pderiv{\phi_m}{r}\right) + \frac{1}{\sin\vartheta}\pderiv{}{\vartheta}\left(\sin\vartheta\pderiv{\phi_m}{\vartheta}\right) + (a_m r)^2\phi_m &= 0\\
	\pderiv{}{r}\left(r^2\pderiv{\psi_{\upvarphi,m}}{r}\right) + \frac{1}{\sin\vartheta}\pderiv{}{\vartheta}\left(\sin\vartheta\pderiv{\psi_{\upvarphi,m}}{\vartheta}\right) + \left[(b_m r)^2-\frac{1}{\sin^2\vartheta}\right]\psi_{\upvarphi,m} &= 0.
\end{align}
Using separation of variables, each of these equations can be reduced to a couple of spherical Bessel and Legendre equations, with the associate Legendre polynomials of zero and first order and spherical Bessel functions as solutions (as described in~\cite{Venas2019e3s}). The solution can be written as\footnote{Note that the Einstein's summation convention will be used throughout this work. Moreover, as in~\cite{Venas2019e3s} the spherical coordinate system $(r,\vartheta,\varphi)$ is used throughout this work.},
\begin{align}
	\phi_m(r,\vartheta) &= \sum_{n=0}^\infty Q_n^{(0)}(\vartheta)A_{m,n}^{(i)}Z_n^{(i)}(\xi_m(r))\label{Eq1:phiSolutionSimplified}\\
	\psi_{\upvarphi,m}(r,\vartheta) &= \sum_{n=0}^\infty Q_n^{(1)}(\vartheta)B_{m,n}^{(i)}Z_n^{(i)}(\eta_m(r)).\label{Eq1:psiSolutionSimplified}
\end{align}
where $A_{m,n}^{(i)}$ and $B_{m,n}^{(i)}$ are coefficients to be found and\footnote{Here, $\besselj_n(x)$ and $\bessely_n(x)$ are the spherical Bessel functions of first and second kind, respectively, $\hankel_n^{(1)}(x)$ is the Hankel function of first kind, and $\legendre_n(x)$ are the Legendre polynomials.}
\begin{align}
	&Z_n^{(1)}(\zeta) = \besselj_n(\zeta),\quad Z_n^{(2)}(\zeta) = \bessely_n(\zeta),\quad Z_n^{(3)}(\zeta) = \hankel_n^{(1)}(\zeta),\quad Q_n^{(j)}(\vartheta) = \deriv[j]{}{\vartheta}\legendre_n(\cos\vartheta),\\
	&\xi_m(r) = a_m r,\quad \eta_m(r) = b_m r.
\end{align}
The derived functions (e.g. the displacement field $u_{\mathrm{r},m}$) can be found in~\Cref{Sec1:derivedFunctions}.

Hysteresis damping~\cite[p. 146]{Skelton1997tao} can be included by modifying Youngs modulus with an addition of an imaginary loss factor for solid domains. However, \cite{Skelton1997tao} assumes the loss factor to be the same in both the longitudinal and transverse wave velocities.  We thus generalize by
\begin{equation}
	c_{\mathrm{s},1,m}\to c_{\mathrm{s},1,m}\sqrt{1-\imag\tilde{\eta}_{1,m}}\quad\text{and}\quad c_{\mathrm{s},2,m}\to c_{\mathrm{s},2,m}\sqrt{1-\imag\tilde{\eta}_{2,m}}
\end{equation}
where $\tilde{\eta}_{1,m}$ and $\tilde{\eta}_{2,m}$ are the loss factor in longitudinal and transverse wave velocities, respectively. For non-viscous fluids this modification is equivalent with 
\begin{equation}
	c_m\to c_m\sqrt{1-\imag\tilde{\eta}_{1,m}}
\end{equation}
but many authors~\cite{Skelton1997tao,Jensen2011coa} do not use the square root for attenuation in non-viscous fluids (i.e., $c_m\to c_m(1-\imag\tilde{\eta}_{1,m})$).
% from which we can compute
% \begin{equation}
% 	E_m = \rho_m c_{\mathrm{s},2,m}^2\frac{3c_{\mathrm{s},1,m}^2-4c_{\mathrm{s},2,m}^2}{c_{\mathrm{s},1,m}^2-c_{\mathrm{s},2,m}^2}.
% \end{equation}

For unbounded domains $\Omega_1$ we impose the Sommerfeld radiation condition~\cite{Sommerfeld1949pde}
\begin{align}\label{Eq1:Sommerfeld}
	\pderiv{\phi_1(\vec{x},\omega)}{r}-\imag a_1 \phi_1(\vec{x},\omega) &= o\left(r^{-1}\right)\\
	\pderiv{\psi_{\upvarphi,1}(\vec{x},\omega)}{r}-\imag b_1 \psi_{\upvarphi,1}(\vec{x},\omega) &= o\left(r^{-1}\right)
\end{align}	
as $r\to\infty$ uniformly in $\hat{\vec{x}}=\frac{\vec{x}}{r}$.

We now introduce prescribed (e.g.~\Cref{Sec1:incidentWave}) incident potential fields $\phi_{\mathrm{inc},m}$ and $\psi_{\upvarphi, \mathrm{inc},m}$ that solves the same equations and has the same derived formulas for the displacement fields, $u_{\mathrm{r},\mathrm{inc},m}$, and stress fields, $\sigma_{\mathrm{rr},\mathrm{inc},m}$, as $\phi_m$ and $\psi_m$, respectively (except, possibly, for the Sommerfeld radiation condition in~\Cref{Eq1:Sommerfeld}). For the derived quantities, such as $u_{\mathrm{r},m}$, we simply add ``inc'' to the subscript (i.e. $u_{\mathrm{r},\mathrm{inc},m}$). For all applications investigated herein, $\psi_{\upvarphi, \mathrm{inc},m}=0$. The coupling conditions (Neumann-to-Neumann, NNBC) for spherical symmetric objects are given by
\begin{align}
	u_{\mathrm{r},m} + u_{\mathrm{r},\mathrm{inc},m} - (u_{\mathrm{r},m+1} + u_{\mathrm{r},\mathrm{inc},m+1}) &= 0,\quad\text{radial displacement boundary condition,}\label{Eq:NNBC_ss_u_r}\\
	u_{\upvartheta,m} + u_{\upvartheta,\mathrm{inc},m} - (u_{\upvartheta,m+1} +u_{\upvartheta,\mathrm{inc},m+1}) &= 0,\quad\text{polar displacement boundary condition,}\label{Eq:NNBC_ss_u_t}\\
	\sigma_{\mathrm{rr},m} + \sigma_{\mathrm{rr},\mathrm{inc},m} - (\sigma_{\mathrm{rr},m+1} + \sigma_{\mathrm{rr},\mathrm{inc},m+1}) &= 0,\quad\text{pressure boundary condition,}\label{Eq:NNBC_ss_sigma_rr}\\
	\sigma_{\mathrm{r}\upvartheta,m} + \sigma_{\mathrm{r}\upvartheta,\mathrm{inc},m} - (\sigma_{\mathrm{r}\upvartheta,m+1} + \sigma_{\mathrm{r}\upvartheta,\mathrm{inc},m+1}) &= 0,\quad\text{traction boundary condition.}\label{Eq:NNBC_ss_sigma_rt}
\end{align}
If $R_M\neq 0$, the following boundary conditions may be implemented on the innermost interface
\begin{align}
	\begin{cases}
	    u_{\mathrm{r},M}+u_{\mathrm{r},\mathrm{inc},M} = 0\\
	    u_{\upvartheta,M}+u_{\mathrm{r},\mathrm{inc},M} = 0,
	\end{cases}\quad&\text{Sound Hard Boundary Condition (SHBC)}\label{Eq:SHBC}\\
	\begin{cases}
	    \sigma_{\mathrm{rr},M} + \sigma_{\mathrm{rr},\mathrm{inc},M} = 0\\
	    \sigma_{\mathrm{r}\upvartheta,M} + \sigma_{\mathrm{r}\upvartheta,\mathrm{inc},M}  = 0,
	\end{cases}\quad&\text{Sound Soft Boundary Condition (SSBC)}\label{Eq:SSBC}\\
	p_M + p_{\mathrm{inc},m}  + \frac{z_M}{\imag k_M \rho_M c_m} \pderiv{(p_M + p_{\mathrm{inc},m})}{r} = 0,\quad&\text{Impedance Boundary Condition (IBC)}\label{Eq:IBC}
\end{align}
with $z_M$ being the impedance~\cite{Ayres1987ars,Lax1948aas}. The impedance condition is used only for non-viscous fluids.

\begin{figure}
	\centering
	\includegraphics{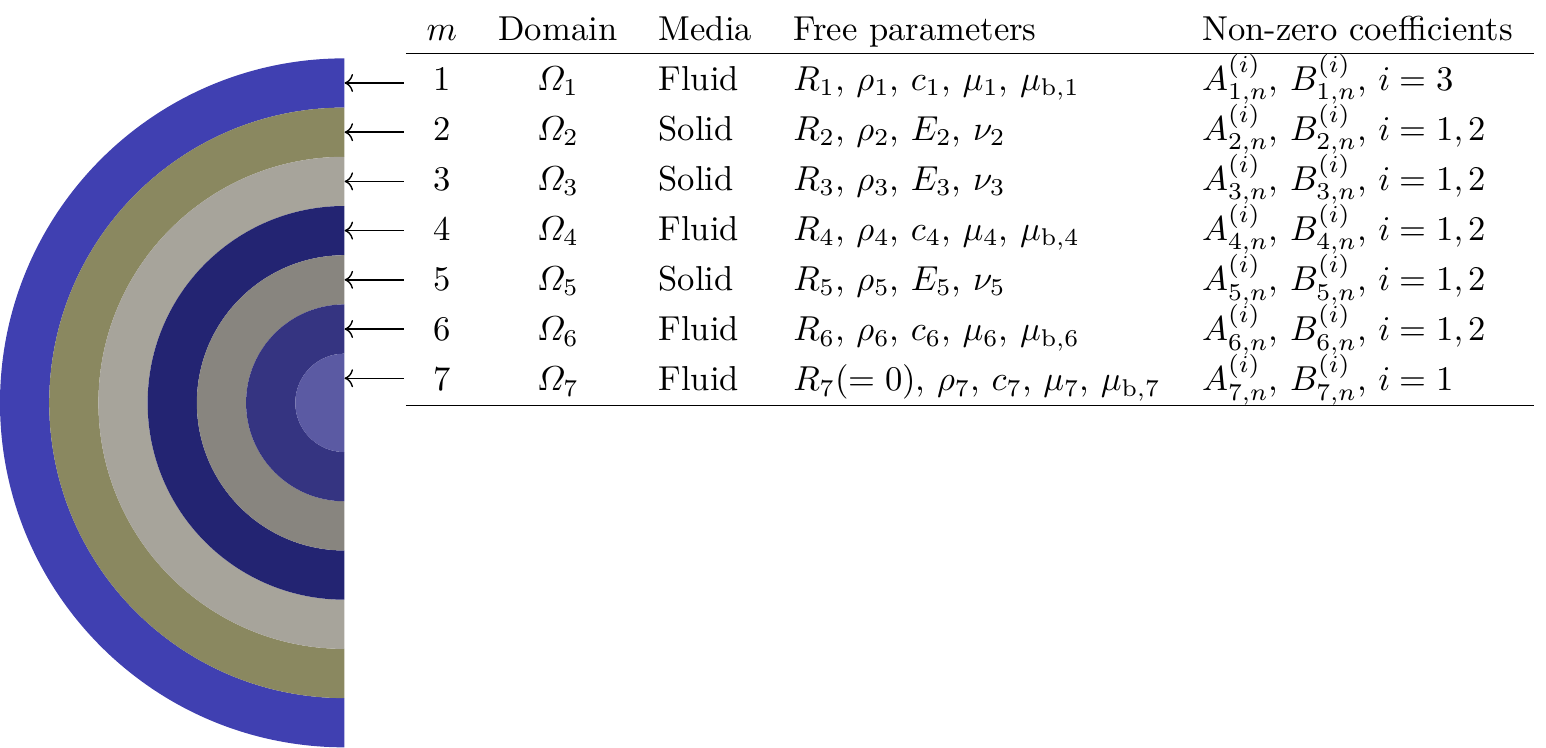}
	\caption{Half of a cross section of a model setup with $M=7$ is described with supporting free parameters and non-zeros coefficients (with the other coefficient being set to zero). The example is similar to the example in~\Cref{Fig1:illustration}, but with equidistant interfaces. Instead of a domain with support at the origin, the boundary conditions \Cref{Eq:SHBC,Eq:SSBC,Eq:IBC} may be used.}
	\label{Fig:e3Dss_setup}
\end{figure}

As the Hankel functions of the first kind satisfies the Sommerfeld condition in~\Cref{Eq1:Sommerfeld}, these will be the radial basis functions for the outermost domain (that is, $A_{1,n}^{(i)}=0$ and $B_{1,n}^{(i)}=0$ for $i=1,2$). For the case $R_M=0$, only the Bessel functions of first kind will constitute a bounded solution and so $A_{M,n}^{(i)}=0$ for $i=2,3$ (see discussion in~\cite{Venas2019e3s}). In all other cases, the Bessel functions of first and second kind are used as the basis for the solution (that is, $A_{m,n}^{(3)}=0$ and $B_{m,n}^{(3)}=0$ for $m>1$). One model setup is shown in~\Cref{Fig:e3Dss_setup}. The linear system of equations (modal equations) are built from boundary conditions at interfaces (starting at the outermost interface, $m=1$).

The non-zero coefficients ($A_{m,n}^{(i)}$ and $B_{m,n}^{(i)}$) are now found by, for each $n$, solving a linear system of equations,
\begin{equation}\label{Eq:LinSysEq}
    \vec{H}_n\vec{C}_n = \vec{D}_n,
\end{equation}
arising from evaluating the boundary conditions~\Cref{Eq:NNBC_ss_u_r,Eq:NNBC_ss_u_t,Eq:NNBC_ss_sigma_rr,Eq:NNBC_ss_sigma_rt,Eq:SHBC,Eq:SSBC,Eq:IBC} at all interfaces (cf.~\cite{Venas2019e3s}).

All solutions are given in terms of spherical Bessel functions which have exponential behavior for large order (as illustrated in~\cite{Venas2019e3s}). This is a problem as one computationally run into ``$0\cdot\infty$'' evaluations (overflow). As we will later see, we can scale the coefficients to include this exponential behavior to mitigate this problem. To find proper scales, a discussion of asymptotic behavior of the Bessel functions is in order.

\section{Scaling Bessel functions}
\label{Sec1:besselLargOrder}
Consider first \textit{cylindrical} Bessel functions of large arguments. As $z\to\infty$ (with $\nu$ fixed) we have the asymptotic expansions~\cite[Eq.~\href{https://dlmf.nist.gov/10.17.E3}{10.17.3}, \href{https://dlmf.nist.gov/10.17.E4}{10.17.4} and \href{https://dlmf.nist.gov/10.17.E4}{10.17.5}]{Olver2022ndl}
\begin{align*}
	\besselJ_{\nu}\left(z\right)&\sim\left(\frac{2}{\pi z}\right)^{\frac{1}{2}}\*\left(\cos q\sum_{k=0}^{\infty}(-1)^{k}\frac{a_{2k}(\nu)}{z^{2k}}-\sin q\sum_{k=0}^{\infty}(-1)^{k}\frac{a_{2k+1}(\nu)}{z^{2k+1}}\right)\\
	\besselY_{\nu}\left(z\right)&\sim\left(\frac{2}{\pi z}\right)^{\frac{1}{2}}\*\left(\sin q\sum_{k=0}^{\infty}(-1)^{k}\frac{a_{2k}(\nu)}{z^{2k}}+\cos q\sum_{k=0}^{\infty}(-1)^{k}\frac{a_{2k+1}(\nu)}{z^{2k+1}}\right)\\
{\Hankel^{(1)}_{\nu}}\left(z\right)&\sim\left(\frac{2}{\pi z}\right)^{\frac{1}{2}}\euler^{\imag q}\sum_{k=0}^{\infty}\imag^{k}\frac{a_{k}(\nu)}{z^{k}}
\end{align*}
where
\begin{equation}
	q = z-\frac{\nu\PI}{2}-\frac{\PI}{4}
\end{equation}
and
\begin{equation}
	a_{k}(\nu)=\frac{(4\nu^{2}-1^{2})(4\nu^{2}-3^{2})\cdots(4\nu^{2}-(2k-1)^{2})}{k!8^{k}}.
\end{equation}
To eliminate exponential behavior in this case, the scaling $\euler^{-|\Im z|}$ can be used for the Bessel functions of first and second kind, and $\euler^{-\imag z}$ for the Hankel function of the first kind.

We consider now the uniform asymptotic expansions for Bessel functions of large order \cite[9.3.38, p. 368]{Abramowitz1965hom}. Define $z(\zeta)$ through
\begin{equation}
	\frac{2}{3}\zeta(z)^{3/2}=\ln\left(\frac{1+\sqrt{1-z^{2}}}{z}\right)-\sqrt{1-z^{2}}.
\end{equation}
Then the uniform asymptotic expansion of the Bessel function of the first kind is given by~\cite[9.3.35-9.3.37, p. 368]{Abramowitz1965hom}
\begin{align}
	\besselJ_{\nu}\left(\nu z\right)&\sim\left(\frac{4\zeta}{1-z^{2}}\right)^{\frac{1}{4}}\*\left(\frac{\mathrm{Ai}\left(\nu^{\frac{2}{3}}\zeta\right)}{\nu^{\frac{1}{3}}}\sum_{k=0}^{\infty}\frac{A_{k}(\zeta)}{\nu^{2k}}+\frac{\mathrm{Ai}'\left(\nu^{\frac{2}{3}}\zeta\right)}{\nu^{\frac{5}{3}}}\sum_{k=0}^{\infty}\frac{B_{k}(\zeta)}{\nu^{2k}}\right)\label{Eq:besselj_asy}\\
	\besselY_{\nu}\left(\nu z\right)&\sim-\left(\frac{4\zeta}{1-z^{2}}\right)^{\frac{1}{4}}\left(\frac{\mathrm{Bi}\left(\nu^{\frac{2}{3}}\zeta\right)}{\nu^{\frac{1}{3}}}\sum_{k=0}^{\infty}\frac{A_{k}(\zeta)}{\nu^{2k}}+\frac{\mathrm{Bi}'\left(\nu^{\frac{2}{3}}\zeta\right)}{\nu^{\frac{5}{3}}}\sum_{k=0}^{\infty}\frac{B_{k}(\zeta)}{\nu^{2k}}\right)\label{Eq:bessely_asy}\\
\Hankel^{(1)}_\nu(\nu z) &\sim 2\euler^{-\PI \imag/3}\left(\frac{4\zeta}{1-z^{2}}\right)^{\frac{1}{4}}\left(\frac{\mathrm{Ai}\left(e^{2\PI \imag/3}\nu^{\frac{2}{3}}\zeta\right)}{\nu^{\frac{1}{3}}}\sum_{k=0}^{\infty}\frac{A_{k}(\zeta)}{\nu^{2k}}+\frac{e^{2\PI \imag/3}\mathrm{Ai}'\left(e^{2\PI \imag/3}\nu^{\frac{2}{3}}\zeta\right)}{\nu^{\frac{5}{3}}}\sum_{k=0}^{\infty}\frac{B_{k}(\zeta)}{\nu^{2k}}\right),\label{Eq:besselh_asy}
\end{align}
as $\nu\to\infty$ and $|\arg z| < \PI$, where
\begin{align*}
	A_{k}(\zeta)&=\sum_{j=0}^{2k}(\tfrac{2}{3}\zeta^{3/2})^{-j}v_{j}U_{2k-j}\left((1-z^{2})^{-\frac{1}{2}}\right)\\
	B_{k}(\zeta)&=-\zeta^{-\frac{1}{2}}\sum_{j=0}^{2k+1}(\tfrac{2}{3}\zeta^{3/2})^{-j}u_{j}U_{2k-j+1}\left((1-z^{2})^{-\frac{1}{2}}\right)
\end{align*}
and $U_k(p)$ are polynomial of degree $3k$ given by $U_0(p)=1$ and
\begin{equation}
	U_{k+1}(p)=\tfrac{1}{2}p^{2}(1-p^{2})U_{k}^{\prime}(p)+\frac{1}{8}\int_{0}^{p}%
(1-5t^{2})U_{k}(t)\mathrm{d}t,
\end{equation}
and finally, starting with $u_0=v_0=1$,
\begin{align*}
	u_{k} &=\frac{(2k+1)(2k+3)(2k+5)\cdots(6k-1)}{216^{k}k!}=\frac{(6k-5)(6k-3)(6k-1%
)}{(2k-1)216k}u_{k-1}\\
	v_{k} &=\frac{6k+1}{1-6k}u_{k}
\end{align*}
for $k=1,2,\dots$.
\begin{remark}\label{Re:transitionRegions}
``Transition regions''~\cite{Olver1952sna} should use separate expansions for when the order is roughly the same size as the magnitude of the argument.
Proper treatment of the transition regions of the asymptotic expansion of the Bessel functions requires usage of another expansion as outlined in~\cite{Olver1952sna}.
\end{remark}

The implementation of the Airy functions in~\cite{Amos1985asp} provides the option of using the scales $\exp(\frac23 y^{3/2})$ and $\exp(-|\frac23\Re{y^{3/2}}|)$ for the Airy functions $\Ai(y)$ and $\Bi(y)$, respectively, in order to eliminate the exponential behavior (same scales are used for the first derivatives). This motivated scales $s_n^{(i)}(x)$, $i=1,2,3$, for the Bessel functions ($\besselj_n(y)$ and $\bessely_n(y)$) and the Hankel function of first kind ($\hankel_n^{(1)}(y)$), respectively, where
\begin{align}
	s_n^{(1)}(x) &= \exp\left\{\frac{2}{3}\nu\zeta(x/\nu)^{\frac{3}{2}}\right\}\label{Eq:s1}\\
	s_n^{(2)}(x) &= \exp\left\{-\left|\Re\left[\frac{2}{3}\nu\zeta(x/\nu)^{\frac{3}{2}}\right]\right|\right\}\label{Eq:s2}\\
	s_n^{(3)}(x) &= \begin{cases}
	\exp\left\{-\left|\Re\left[\frac{2}{3}\nu\zeta(x/\nu)^{\frac{3}{2}}\right]\right|\right\} & \Im x < 0\\
	\exp\left\{-\frac{2}{3}\nu\zeta(x/\nu)^{\frac{3}{2}}\right\} & \text{otherwise.}
	\end{cases}\label{Eq:s3}
\end{align}
with $\nu=n+1/2$. 

\begin{figure}
	\centering
	\begin{subfigure}[t]{0.38\textwidth}
		\includegraphics[width=\textwidth]{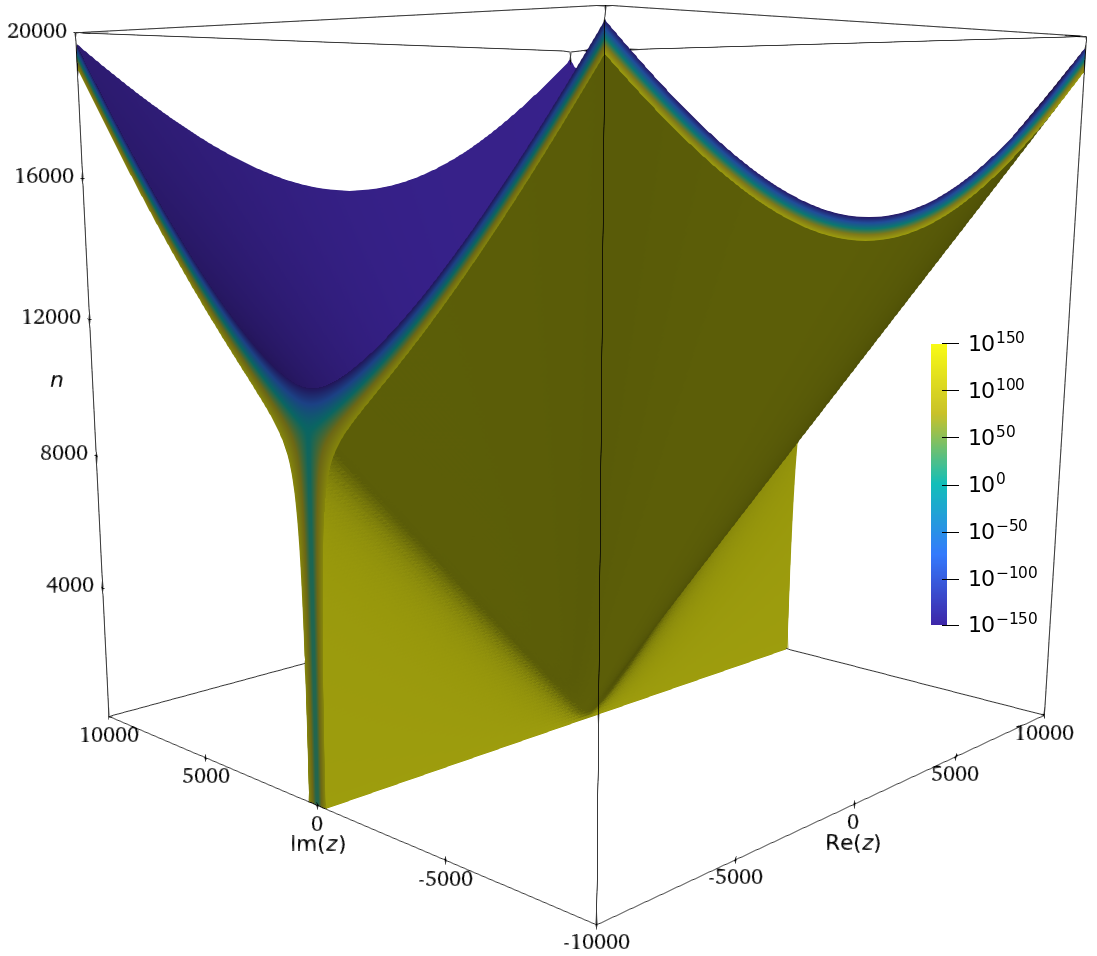}
		\caption{Plot of $\left|Z_n^{(1)}(z)\right|$.}
	\end{subfigure}
	\hspace{1cm}
	\begin{subfigure}[t]{0.38\textwidth}
		\includegraphics[width=\textwidth]{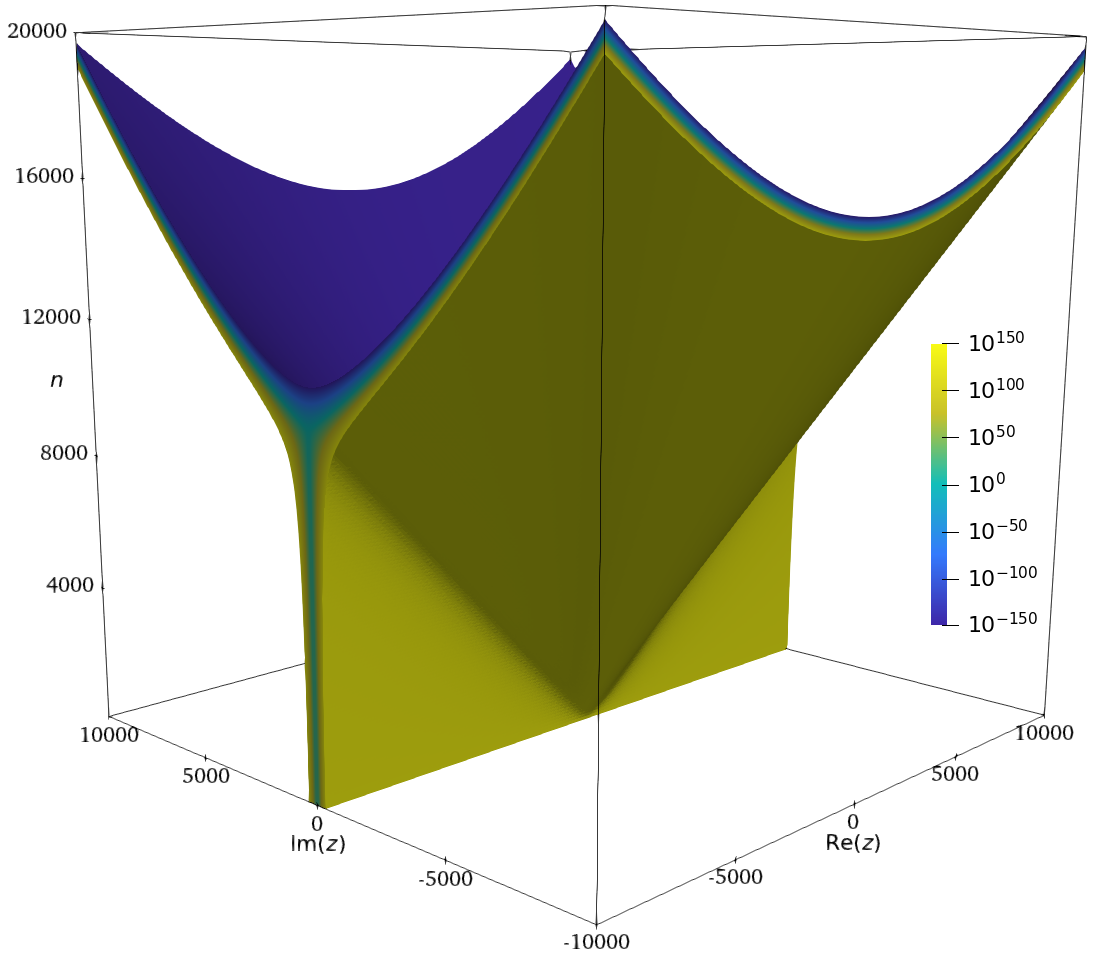}
		\caption{Plot of $\left|s_n^{(1)}(z)\right|^{-1}$.}
	\end{subfigure}
	\par\bigskip
	\begin{subfigure}[t]{0.38\textwidth}
		\includegraphics[width=\textwidth]{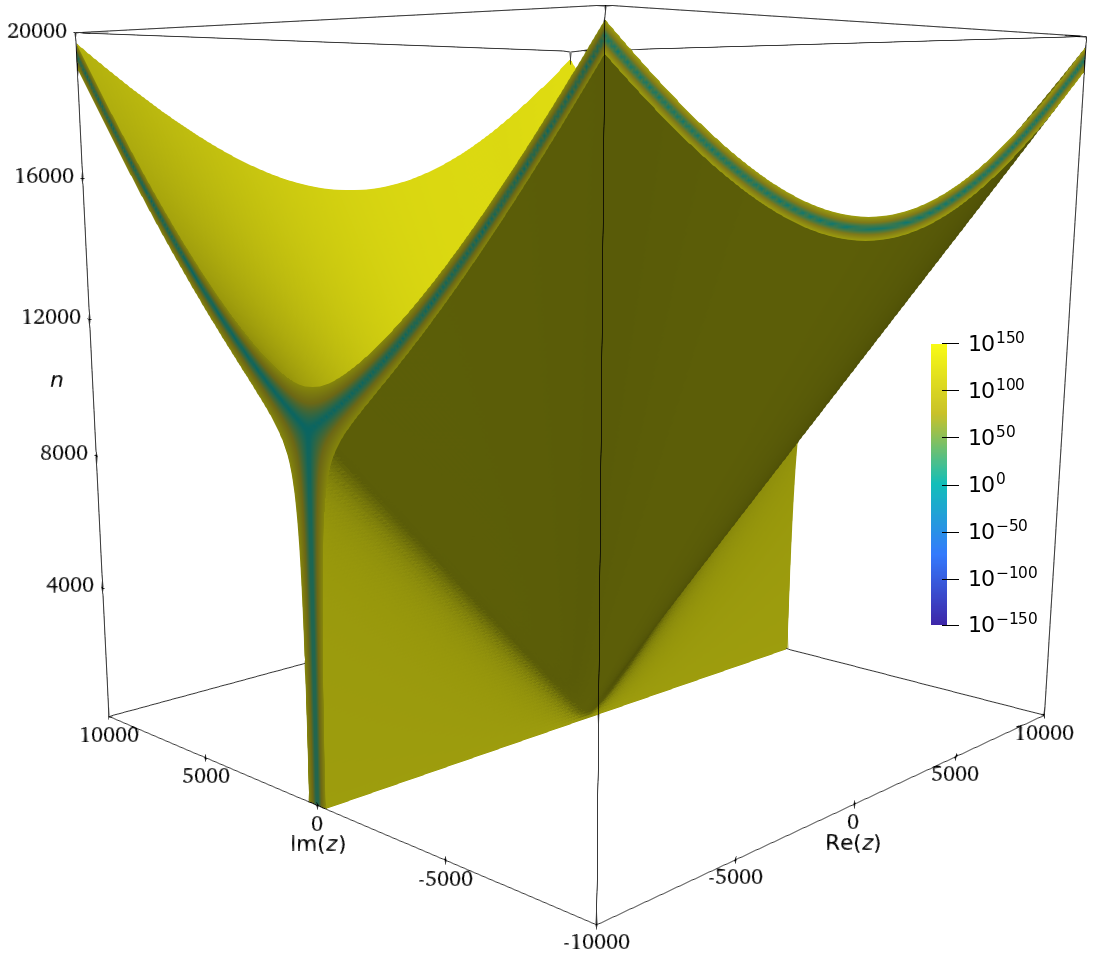}
		\caption{Plot of $\left|Z_n^{(2)}(z)\right|$.}
	\end{subfigure}
	\hspace{1cm}
	\begin{subfigure}[t]{0.38\textwidth}
		\includegraphics[width=\textwidth]{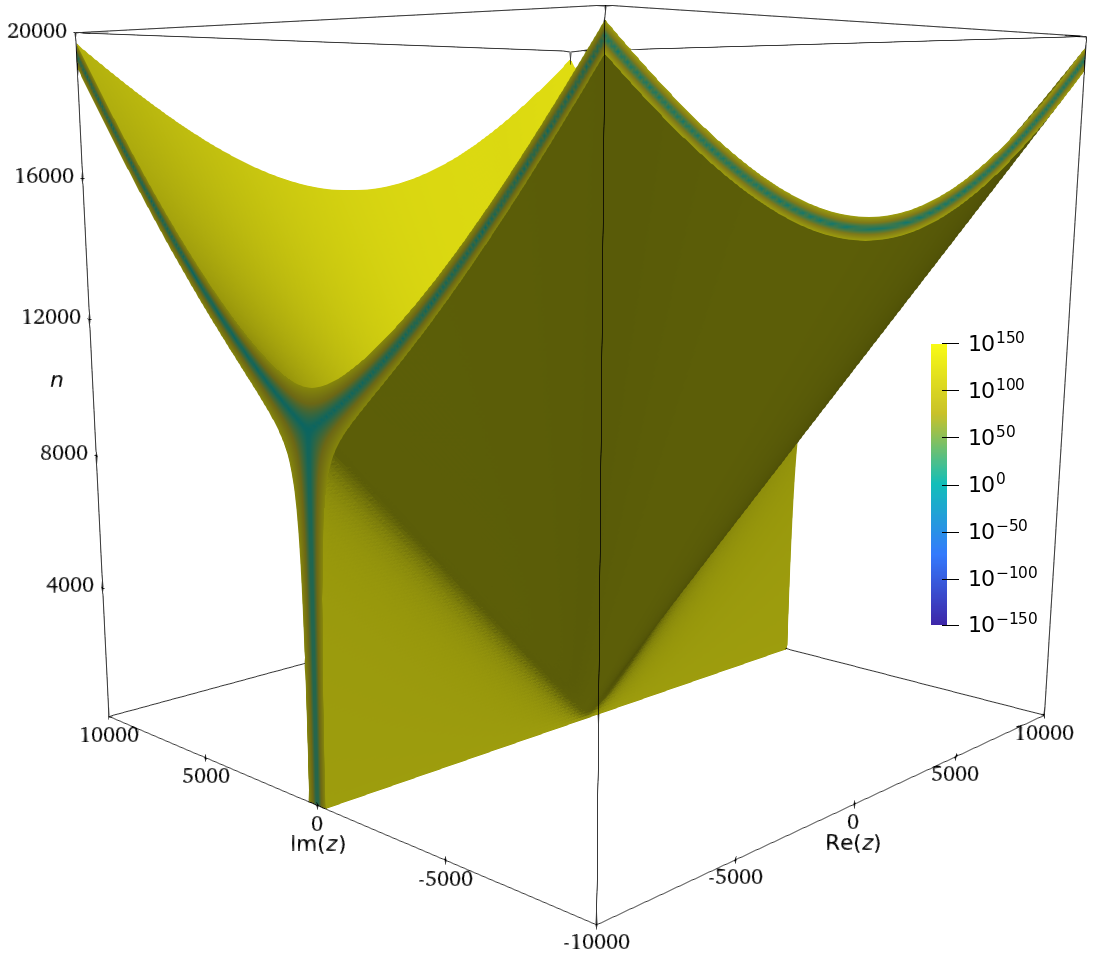}
		\caption{Plot of $\left|s_n^{(2)}(z)\right|^{-1}$.}
	\end{subfigure}
	\par\bigskip
	\begin{subfigure}[t]{0.38\textwidth}
		\includegraphics[width=\textwidth]{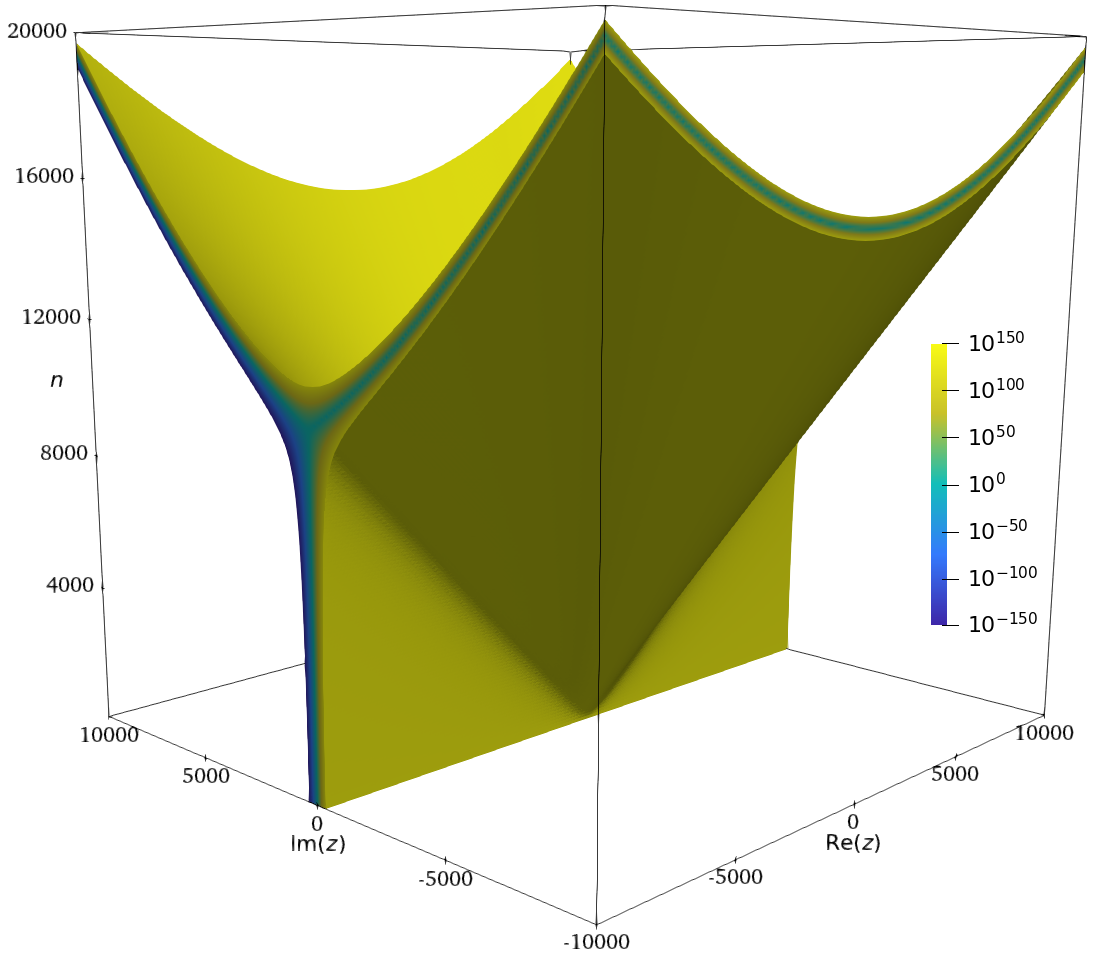}
		\caption{Plot of $\left|Z_n^{(3)}(z)\right|$.}
	\end{subfigure}
	\hspace{1cm}
	\begin{subfigure}[t]{0.38\textwidth}
		\includegraphics[width=\textwidth]{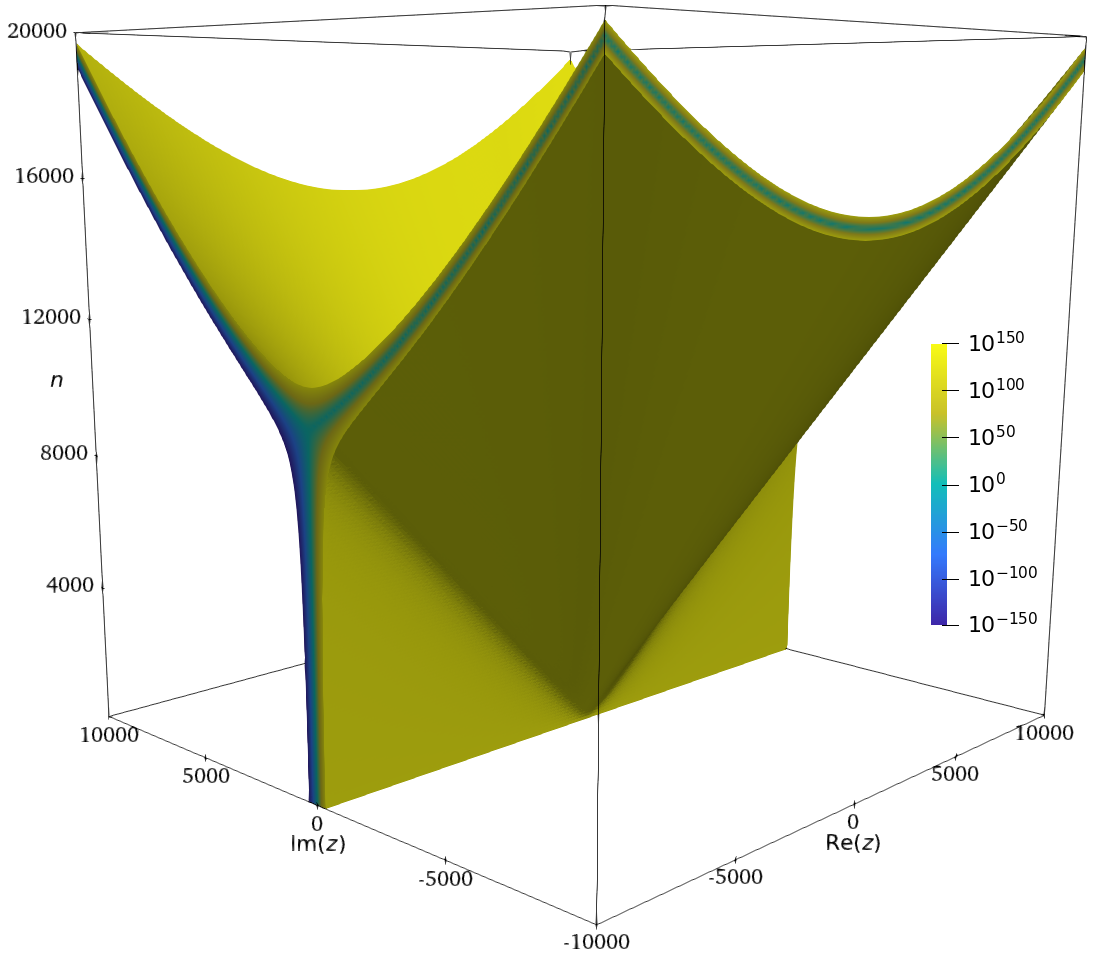}
		\caption{Plot of $\left|s_n^{(3)}(z)\right|^{-1}$.}
	\end{subfigure}
	\caption{Plots of the magnitude of the Bessel functions and their scales. The plots are obtained by sampling the coordinates $(\Re z, \Im z, n)$ on a $200\times 200\times 200$ grid where the real part and the imaginary part ranges from \num{-e4} to \num{e4}, and the order from \num{0} to \num{2e4}. The solution is then clipped for values outside the interval $[10^{-150}, 10^{150}]$. The visualizations are done using \href{https://www.paraview.org}{Paraview}.}
	\label{Fig:besselMag}
\end{figure}

In~\Cref{Fig:besselMag} the magnitude of the Bessel functions and their scales are compared on the domain 
\begin{equation}
	\mathcal{D} = \left\{(n,z)\in\N\times\C\st n\in\left[0,\num{2e4}\right],\quad \Re z\in\left[-10^4,10^4\right],\quad \Im z\in\left[-10^4,10^4\right]\right\}.
\end{equation}
The plots are visually indistinguishable validating the correct scales $s_n^{(i)}(z)$ for the Bessel functions $Z_n^{(i)}(z)$. The usage of \Cref{Eq:besselj_asy,Eq:bessely_asy,Eq:besselh_asy} will be limited to the inside of the ``cup''\footnote{Defined by
\begin{equation*}
	\mathcal{C} = \left\{z\in\C\st \left|s_n^{(1)}(z)\right|^{-1} < \sqrt{\realmin},\quad \nu > \nu_{\mathrm{a}} \right\}
\end{equation*}
where $\realmin$ is the minimum floating-point number for the precision used. For double precision we have
\begin{align*}
	\realmin &=  2^{-1022}                 \approx 2.2250738585072013830902327173324 \cdot 10^{-308}\\
	\realmax &=  (2-2^{-52})\cdot 2^{1023} \approx 1.7976931348623157081452742373170 \cdot 10^{308}.
\end{align*} 
Throughout this work $\nu_{\mathrm{a}}=100$ has been used. 
}
in these plots as $n$ must be larger than the magnitude of $z$ for these formula to apply. However, the scales motivated by the same formulas works remarkably well in other parts of the domain as well. Further evidence for this can be obtained using high precision toolboxes like the multiprecision computing toolbox \href{https://www.advanpix.com/}{Advanpix}. A test was performed (using 100 digits precision) in the same domain used in~\Cref{Fig:besselMag} which resulted in the bounds $\num{2.718e-06} \lessapprox |Z_n^{(i)}(z)/s_n^{(i)}(z)| \lessapprox \num{6.657e-02}$, $i=1,2,3$, for $(n,z)\in\mathcal{D}$ using $41\times 41\times 41$ uniformly sampled points (for reference: $\max_{i,n,z}|Z_n^{(i)}(z)| \approx \num{9.341e+29375}$, $i=1,2,3$, $(n,z)\in\mathcal{D}$). The ``$0/0$'' evaluations are here discarded.

\section{Computationally stable solutions}
\label{Sec:stableSol}
In the following we transform the expressions in~\cite{Venas2019e3s} to mitigate computational instability based on the discussion in previous section. The scales are always applied on the final formulas in~\cite{Venas2019e3s} (that is, after differentiation and manipulations). Define the intermediate radii by
\begin{equation}
	\tilde{R}_m = \begin{cases}
		R_1 & m = 1\\
		R_{M-1} & m = M\quad\text{and}\quad R_M=0\\
		(R_m-R_{m-1})/2 & \text{otherwise}.
		\end{cases}		
\end{equation}
With
\begin{align*}
	\tilde{Z}_n^{(i)}(x) &= s_n^{(i)}(x) Z_n^{(i)}(x),\qquad i=1,3\\
	\tilde{A}_{m,n}^{(i)} &= A_{m,n}^{(i)}\left[s_n^{(i)}(\xi_m(\tilde{R}_m))\right]^{-1},\qquad i=1,2\\
	\tilde{B}_{m,n}^{(i)} &= B_{m,n}^{(i)}\left[s_n^{(i)}(\eta_m(\tilde{R}_m))\right]^{-1},\qquad i=1,2\\
	\tilde{C}_{m,n}^{(i)} &= C_{m,n}^{(i)}\left[s_n^{(i)}(\zeta_m(\tilde{R}_m))\right]^{-1},\qquad i=1,3
\end{align*}
we can write the solution as
\begin{align}
	\phi_m &= \sum_{n=0}^\infty Q_n^{(0)}(\vartheta)\tilde{A}_{m,n}^{(i)}w_n^{(i)}(\xi_m(\tilde{R}_m),\xi_m(r))\tilde{Z}_n^{(i)}(\xi_m(r))\\
	\psi_{\upvarphi,m} &= \sum_{n=0}^\infty Q_n^{(1)}(\vartheta)\tilde{B}_{m,n}^{(i)}w_n^{(i)}(\eta_m(\tilde{R}_m),\eta_m(r))\tilde{Z}_n^{(i)}(\eta_m(r))
\end{align}
where
\begin{equation}
	w_n^{(i)}(x,y) = \frac{s_n^{(i)}(x)}{s_n^{(i)}(y)}.
\end{equation}
\begin{remark}\label{Re:overflow}
Computationally, it is here important to evaluate the combined arguments for the exponential factors in the numerator and the denominator before applying the exponential function (in order to avoid overflow).
\end{remark}
\begin{remark}\label{Re:overflow2}
For large $n$ and $|x-y|$ even the combined evaluations (\Cref{Re:overflow}) will yield overflow, and the strategy presented herein only extends the range of parameters at which the solution may be computationally evaluated (for a given floating point precision).
\end{remark}

For the displacement field we correspondingly have (from~\Cref{Eq1:u_rgen,Eq1:u_tgen})
\begin{equation}\
	u_{\mathrm{r},m} = \frac{1}{r}\sum_{n=0}^\infty Q_n^{(0)}(\vartheta)\left[\tilde{A}_{m,n}^{(i)}w_n^{(i)}(\xi_m(\tilde{R}_m),\xi_m(r))\tilde{S}_{1,n}^{(i)}(\xi_m(r))+\tilde{B}_{m,n}^{(i)}w_n^{(i)}(\eta_m(\tilde{R}_m),\eta_m(r))\tilde{T}_{1,n}^{(i)}(\eta_m(r))\right]
\end{equation}
and
\begin{equation}
	u_{\upvartheta,m} = \frac{1}{r}\sum_{n=0}^\infty Q_n^{(1)}(\vartheta)\left[\tilde{A}_{m,n}^{(i)}w_n^{(i)}(\xi_m(\tilde{R}_m),\xi_m(r))\tilde{S}_{2,n}^{(i)}(\xi_m(r))+\tilde{B}_{m,n}^{(i)}w_n^{(i)}(\eta_m(\tilde{R}_m),\eta_m(r))\tilde{T}_{2,n}^{(i)}(\eta_m(r))\right]
\end{equation}
where
\begin{align*}
	\tilde{S}_{1,n}^{(i)}(\xi) &= n\tilde{Z}_n^{(i)}(\xi)-\xi g_n^{(i)}(\xi)\tilde{Z}_{n+1}^{(i)}(\xi)\\ 
	\tilde{T}_{1,n}^{(i)}(\eta) &= -n(n+1)\tilde{Z}_n^{(i)}(\eta)\\
	\tilde{S}_{2,n}^{(i)}(\xi) &= \tilde{Z}_n^{(i)}(\xi)\\
	\tilde{T}_{2,n}^{(i)}(\eta) &= -(n+1)\tilde{Z}_n^{(i)}(\eta) + g_n^{(i)}(\eta)\eta \tilde{Z}_{n+1}^{(i)}(\eta).
\end{align*}
and (\Cref{Re:overflow} is also important here)
\begin{align}\label{Eq:gDef}
	g_n^{(i)}(x) = \frac{s_n^{(i)}(x)}{s_{n+1}^{(i)}(x)}.
\end{align}
\begin{remark}\label{Re:g}
For a fixed $x$ one can show that $g_n^{(1)}(x) \sim \frac{x}{2n}$, $g_n^{(2)}(x) \sim \frac{2n}{|x|}$ and $g_n^{(3)}(x) \sim \frac{2n}{x}$, as $n\to\infty$, such that $g_n^{(i)}$ should not be considered to have exponential behavior in this context.
\end{remark}

The linear system of equation in~\Cref{Eq:LinSysEq} is now modified to be
\begin{equation}
    \tilde{\vec{H}}_n\tilde{\vec{C}}_n=\vec{D}_n,
\end{equation}
where $\tilde{\vec{H}}_n$ and $\tilde{\vec{C}}_n$ are modifications of $\vec{H}_n$ and $\vec{C}_n$ taking into account the scaling functions $w_n^{(i)}(x,y)$ and $s_n^{(i)}(x)$, respectively.

\begin{figure}
	\centering
	\begin{subfigure}[t]{0.48\textwidth}
		\includegraphics[scale=0.95]{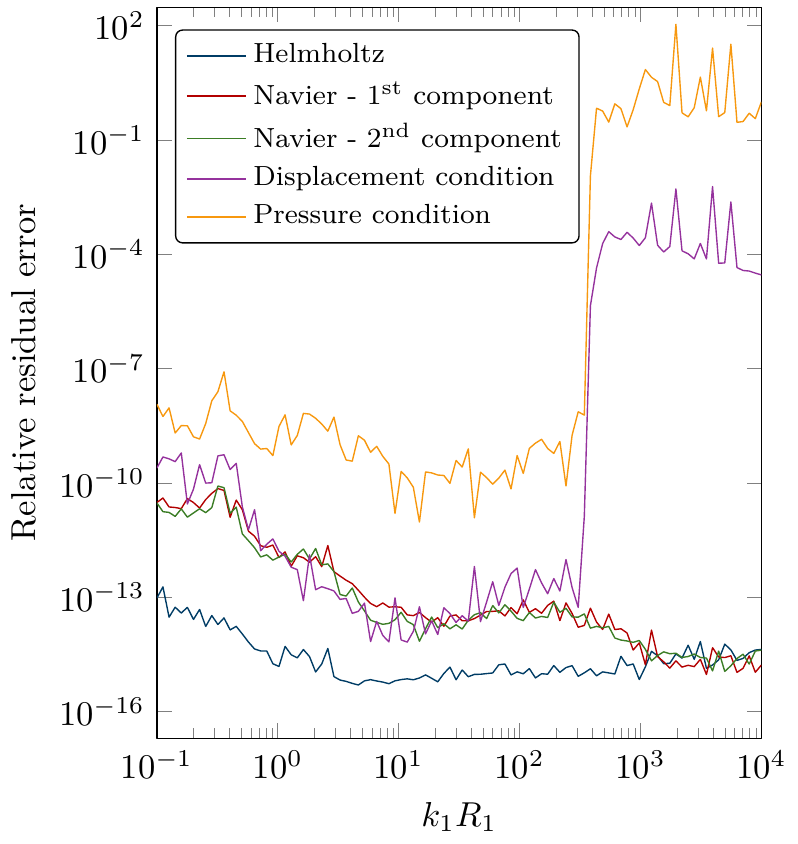}
		\caption{Without scaling using double precision}
	\end{subfigure}
	~
	\begin{subfigure}[t]{0.48\textwidth}
		\includegraphics[scale=0.95]{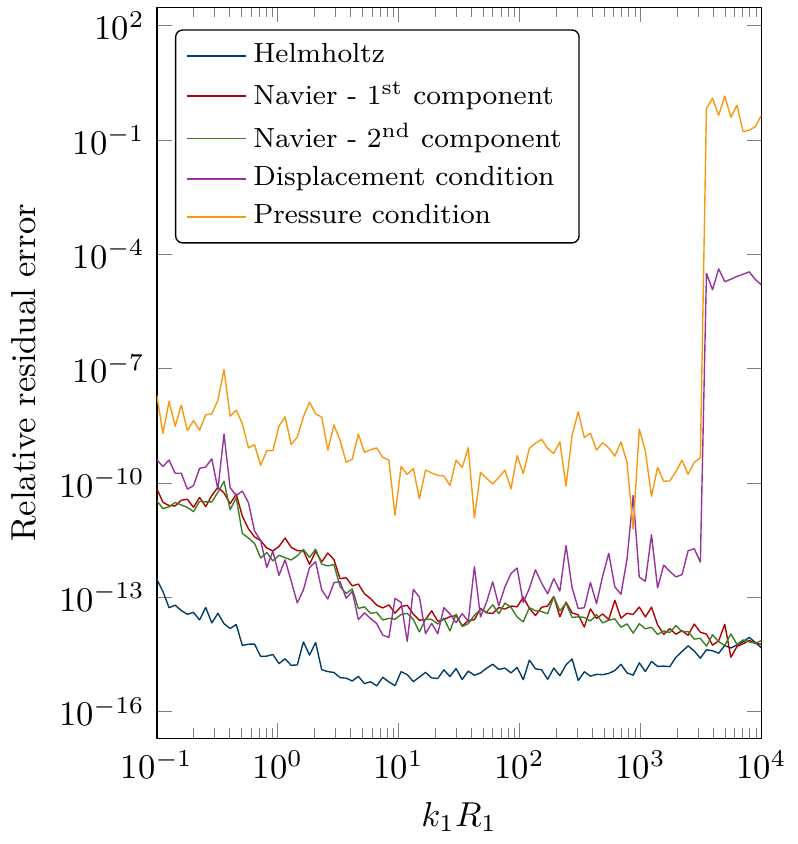}
		\caption{With scaling using double precision}
	\end{subfigure}
	\par\bigskip
	\begin{subfigure}[t]{0.48\textwidth}
		\includegraphics[scale=0.95]{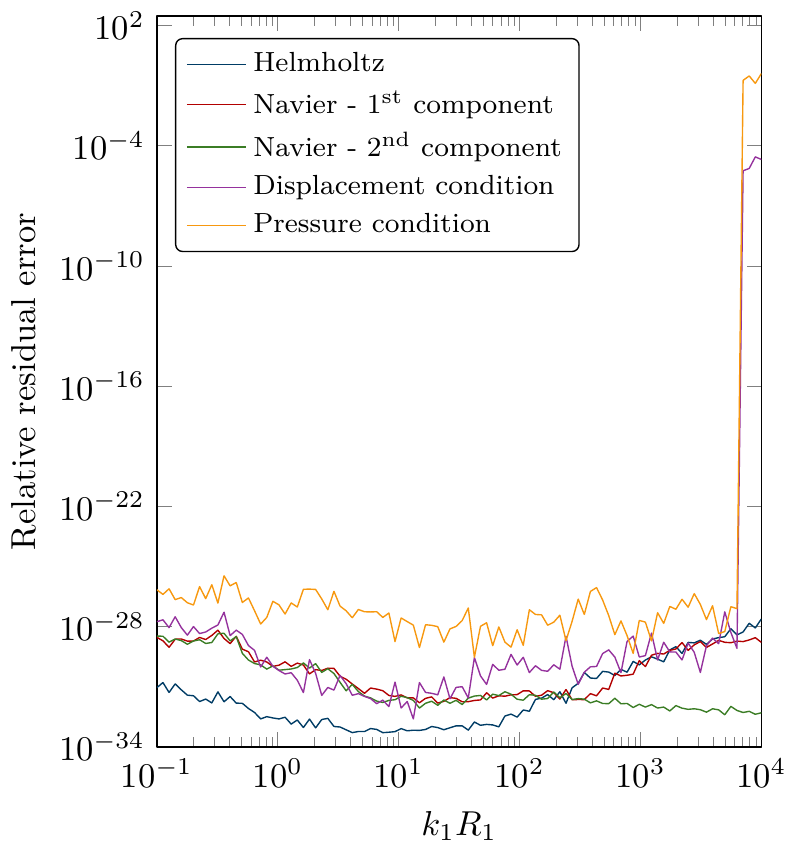}
		\caption{Without scaling using quadruple precision}
	\end{subfigure}
	~
	\begin{subfigure}[t]{0.48\textwidth}
		\includegraphics[scale=0.95]{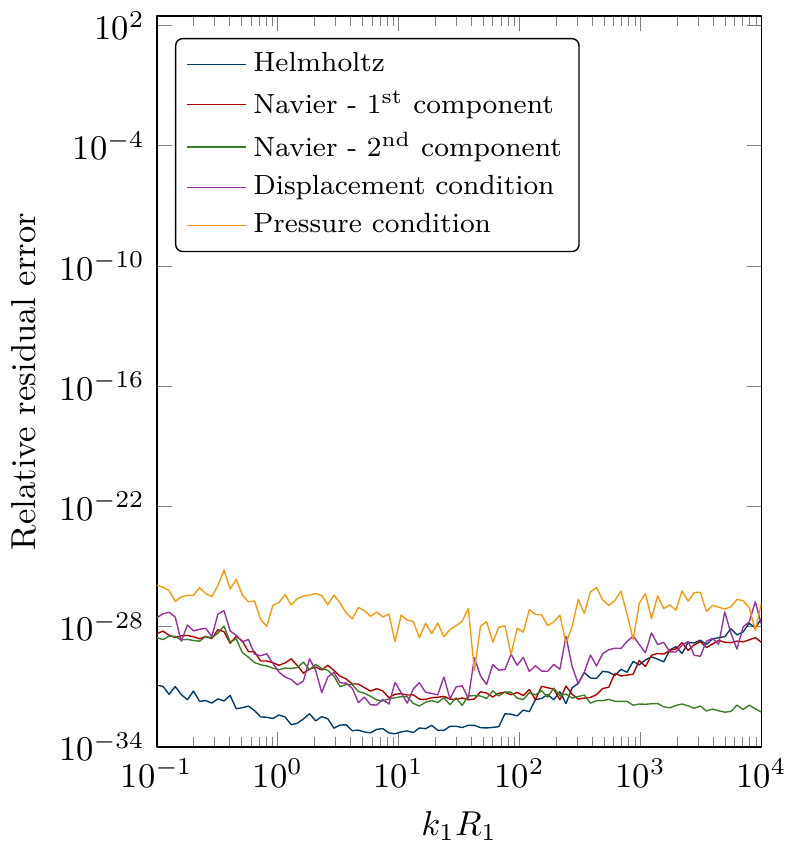}
		\caption{With scaling using quadruple precision}
	    \label{Fig:errorsS123_SSBC_4}
	\end{subfigure}
	\caption{\textbf{S135 benchmark}: Residual errors for the governing equations and boundary conditions. The quadruple calculations were enabled through \href{https://www.advanpix.com/}{Advanpix} in \MATLAB.}
	\label{Fig:errorsS123_SSBC}
\end{figure}

In~\Cref{Fig:errorsS123_SSBC} the setup used in figure 6 in~\cite{Venas2019e3s} (with the S135 model with SHBC on the inner sphere and parameters in~\Cref{Tab:S135}) is used to illustrate the improvement of this scaling.
\begin{table}
	\centering
	\caption{\textbf{S135 benchmark} Parameters for the examples in~\Cref{Fig:errorsS123_SSBC}.}
	\label{Tab:S135}
	\begin{tabular}{l l}
		\toprule
		Parameter & Description\\
		\midrule
		$R_1 = \SI{5}{m}$ & Inner radius of outer fluid (domain 1)\\
		$\rho_1 = \SI{1000}{kg.m^{-3}}$ & Density of outer fluid (domain 1)\\
		$c_1 = \SI{1500}{m.s^{-1}}$ & Speed of sound in outer fluid (domain 1)\\
		$R_2 = \SI{4.992}{m}$ & Inner radius of outer shell (domain 2)\\
		$\rho_2= \SI{7850}{kg.m^{-3}}$ & Density of outer shell (domain 2)\\
		$E_2 = \SI{210e9}{Pa}$ & Young's modulus of outer shell (domain 2)\\
		$\nu_2 = 0.3$ & Poisson's ratio of outer shell (domain 2)\\
		$R_3 = \SI{3}{m}$ & Inner radius of intermediate fluid (domain 3)\\
		$\rho_3 = \SI{1000}{kg.m^{-3}}$ & Density of intermediate fluid (domain 3)\\
		$c_3 = \SI{1500}{m.s^{-1}}$ & Speed of sound in intermediate fluid (domain 3)\\
		$R_4 = \SI{2.98}{m}$ & Inner radius of intermediate solid shell (domain 4)\\
		$\rho_4= \SI{7850}{kg.m^{-3}}$ & Density of intermediate solid shell (domain 4)\\
		$E_4 = \SI{210e9}{Pa}$ & Young's modulus of intermediate solid shell (domain 4)\\
		$\nu_4 = 0.3$ & Poisson's ratio of intermediate solid shell (domain 4)\\
		$R_5 = \SI{1}{m}$ & Inner radius of inner fluid (domain 5)\\
		$\rho_5 = \SI{1000}{kg.m^{-3}}$ & Density of inner fluid (domain 5)\\
		$c_5 = \SI{1500}{m.s^{-1}}$ & Speed of sound of inner fluid (domain 5)\\
		\bottomrule
	\end{tabular}
\end{table}
The cut-off at which machine epsilon precision is no longer obtained\footnote{By this we mean when the ratio between term $N_\varepsilon$ and its sum in absolute value is less than machine epsilon precision.} is increased from $k_1R_1\approx 107$ to $k_1R_1\approx 614$. This limit is extended by roughly a factor 10 if quadruple-precision is used (which increases $\realmax$).
At the end however, exponential functions must still be evaluated which still limits computational evaluations for high enough frequencies. To extend the computational frequency range one needs a representation with more bits for the exponent than the standard 8-bit representation for the exponent for double precision. In~\Cref{Fig:errorsS123_SSBC_4} quadruple precision is used and covers the frequency range for most applications.
\section{Numerical examples} 
\label{Sec1:resultsDisc}
To give further evidence for the correctness of the implemented code, comparison to existing benchmark solutions by Sage~\cite{Sage1979mri}, Skelton~\cite{Skelton1997tao}, Hetmaniuk~\cite{Hetmaniuk2012raa}, and Ayres~\cite{Ayres1987ars}, will be presented.

It is customary to present results in the \textit{far-field}. For the scattered pressure $p_1$, it is defined by
\begin{equation}
	p_0(\hat{\vec{x}},\omega) =  r \euler^{-\imag a_1 r}p_1(\vec{x},\omega),\quad r = |\vec{x}| \to \infty,
\end{equation}
with $\hat{\vec{x}} = \vec{x}/|\vec{x}|$.

From the far-field pattern, the \textit{target strength}, $\TS$, can be computed. It is defined by
\begin{equation}\label{Eq1:TS}
	\TS = 20\log_{10}\left(\frac{|p_0(\hat{\vec{x}},\omega)|}{|P_{\mathrm{inc}}(\omega)|}\right)
\end{equation}
where $P_{\mathrm{inc}}$ is the amplitude of the incident wave at the geometric center of the scatterer (i.e., the origin).

Unless stated otherwise, the direction of the incident wave will be given by the vector $\vec{d}_{\mathrm{s}} = [0,0,1]^\transpose$.

\subsection{Sage benchmark problem} 
\begin{figure}
	\centering
	\begin{subfigure}[t]{\textwidth}
		\centering
		\includegraphics{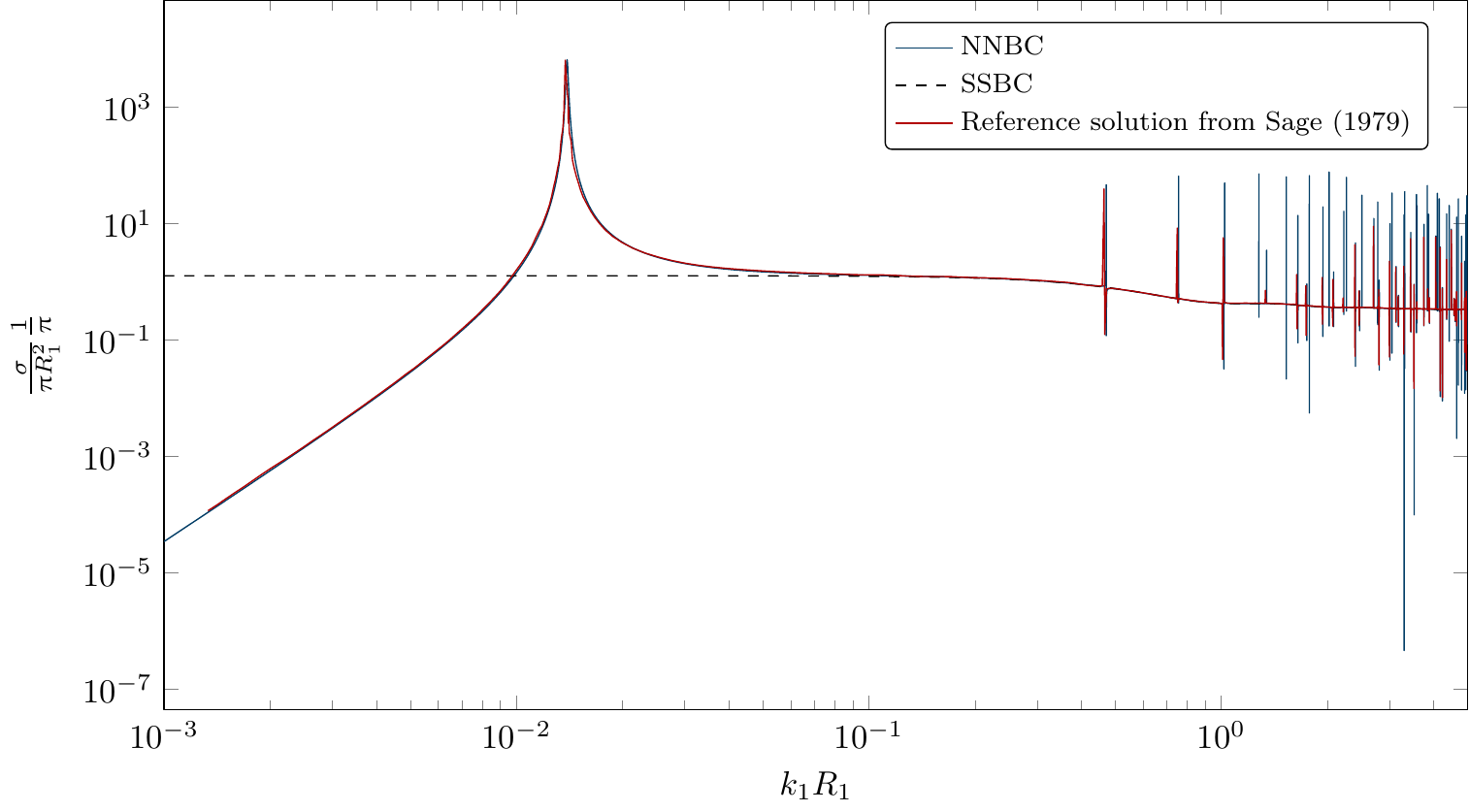}
	\end{subfigure}
	\par\bigskip
	\begin{subfigure}[t]{\textwidth}
		\centering
		\includegraphics{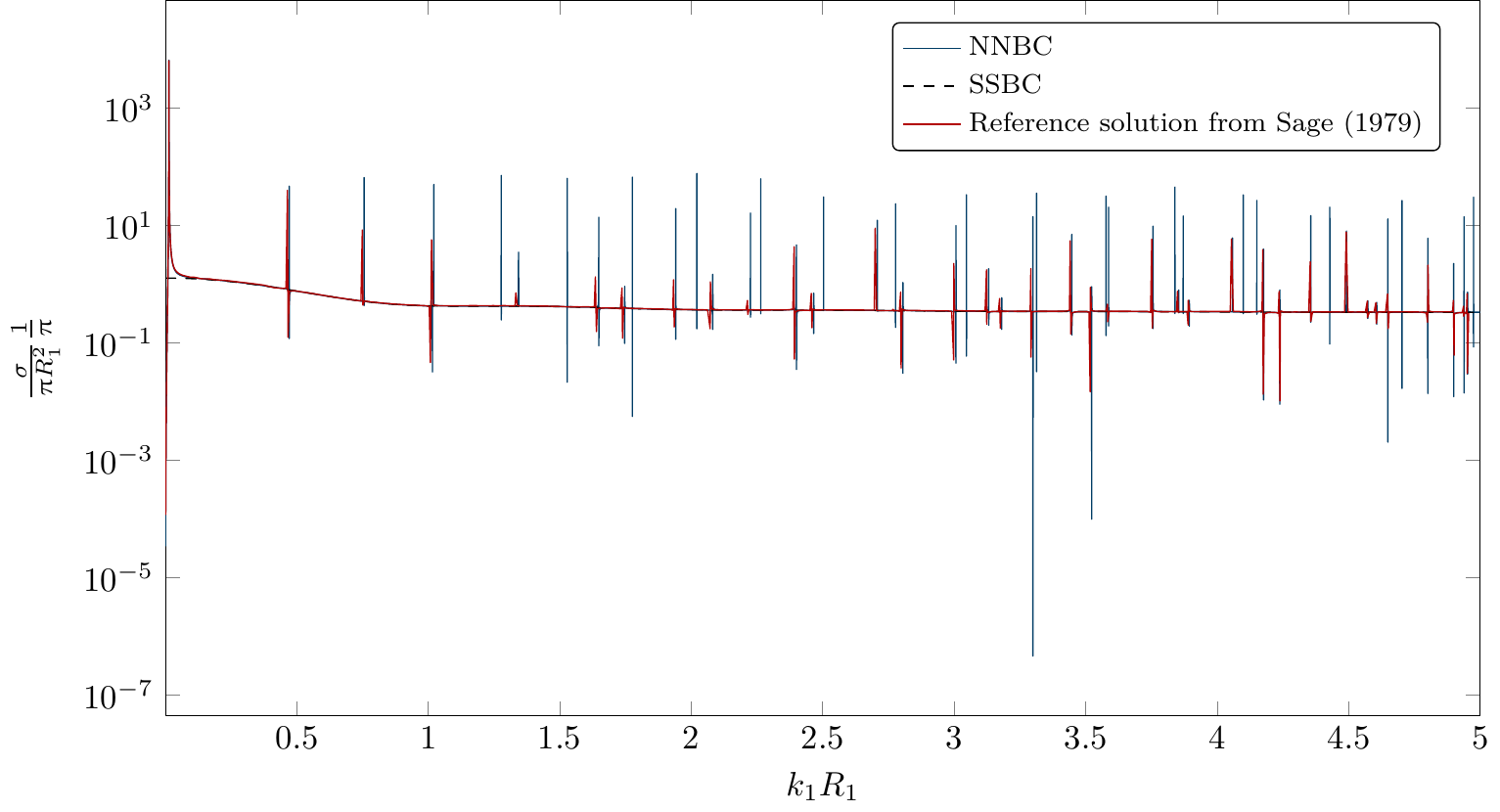}
	\end{subfigure}
	\caption{\textbf{Sage benchmark problem}: Sonar cross section of an air bubble in water.}
	\label{Fig:Sage}
\end{figure}

Sage~\cite{Sage1979mri} considers an air bubble and uses the parameters in \Cref{Tab1:Sage}.
\begin{table}
	\centering
	\caption{\textbf{Sage parameters:} Parameters for the examples in Fig. 1 and Fig. 4 in \cite{Sage1979mri}.}
	\label{Tab1:Sage}
	\begin{tabular}{l l}
		\toprule
		Parameter & Description\\
		\midrule
		$\rho_{1} = \SI{1025}{kg.m^{-3}}$ & Density of water\\
		$c_{1} = \SI{1531}{m.s^{-1}}$ & Speed of sound in water at $\SI{25}{\celsius}$\\
		$\rho_{2} = \SI{1.293}{kg.m^{-3}}$ & Density of air\\
		$c_{2} = \SI{346.2}{m.s^{-1}}$ & Speed of sound in air at $\SI{25}{\celsius}$\\
		\bottomrule
	\end{tabular}
\end{table}
Sage defines the cross-section area by
\begin{equation}
	\sigma = 4\PI \frac{|p_0|^2}{|P_{\mathrm{inc}}|^2}.
\end{equation}
In \Cref{Fig:Sage} the present work treating the interface as fluid-fluid interaction (NNBC) and treating the interface as sound soft boundary condition (SSBC) are compared with the data from Fig. 1 and Fig. 2 in~\cite{Sage1979mri}\footnote{The discrepancies probably comes from the fact that the data set is collected by the software \href{https://automeris.io/WebPlotDigitizer/}{WebPlotDigitizer} where a digital scan of Fig. 1 and Fig. 4 in~\cite{Sage1979mri} has been made.}. Moreover, the spectrum has been sampled closely, revealing small (less significant) eigenmodes not shown by Sage. It seems as if the $y$-axis in~\cite{Sage1979mri} is scaled by $1/\PI$, and so this is also done here. A good agreement is here observed.

\subsection{Skelton benchmark problem} 
\begin{figure}
	\includegraphics{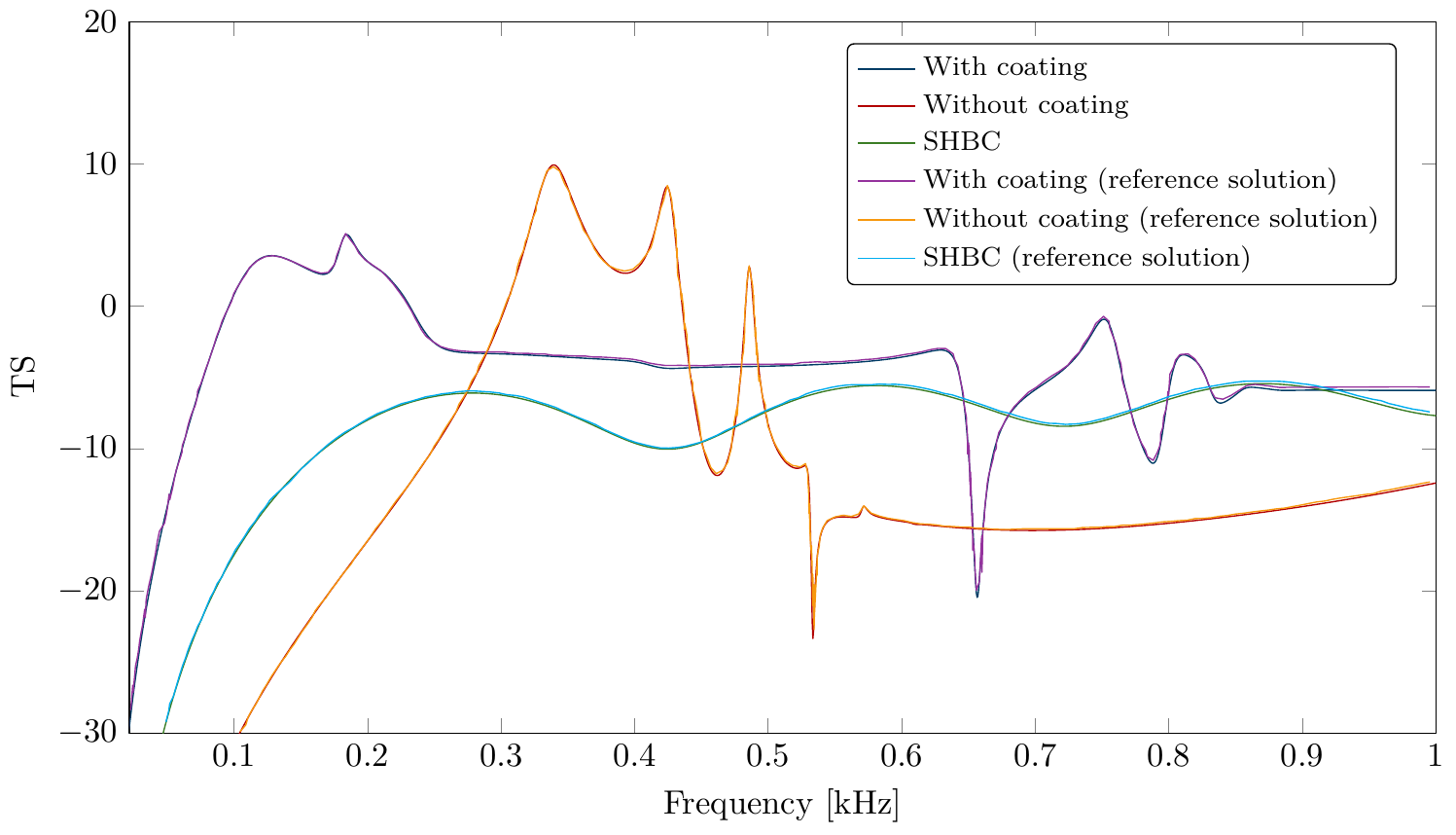}
	\caption{\textbf{Skelton benchmark problem}: An empty steel shell is considered with and without coating. The sound hard sphere is also considered. All three cases are compared to reference solutions from~\cite{Skelton1997tao}.}
	\label{Fig:Skelton}
\end{figure}

\begin{figure}
	\centering
	\begin{subfigure}[t]{\textwidth}
		\centering
		\includegraphics{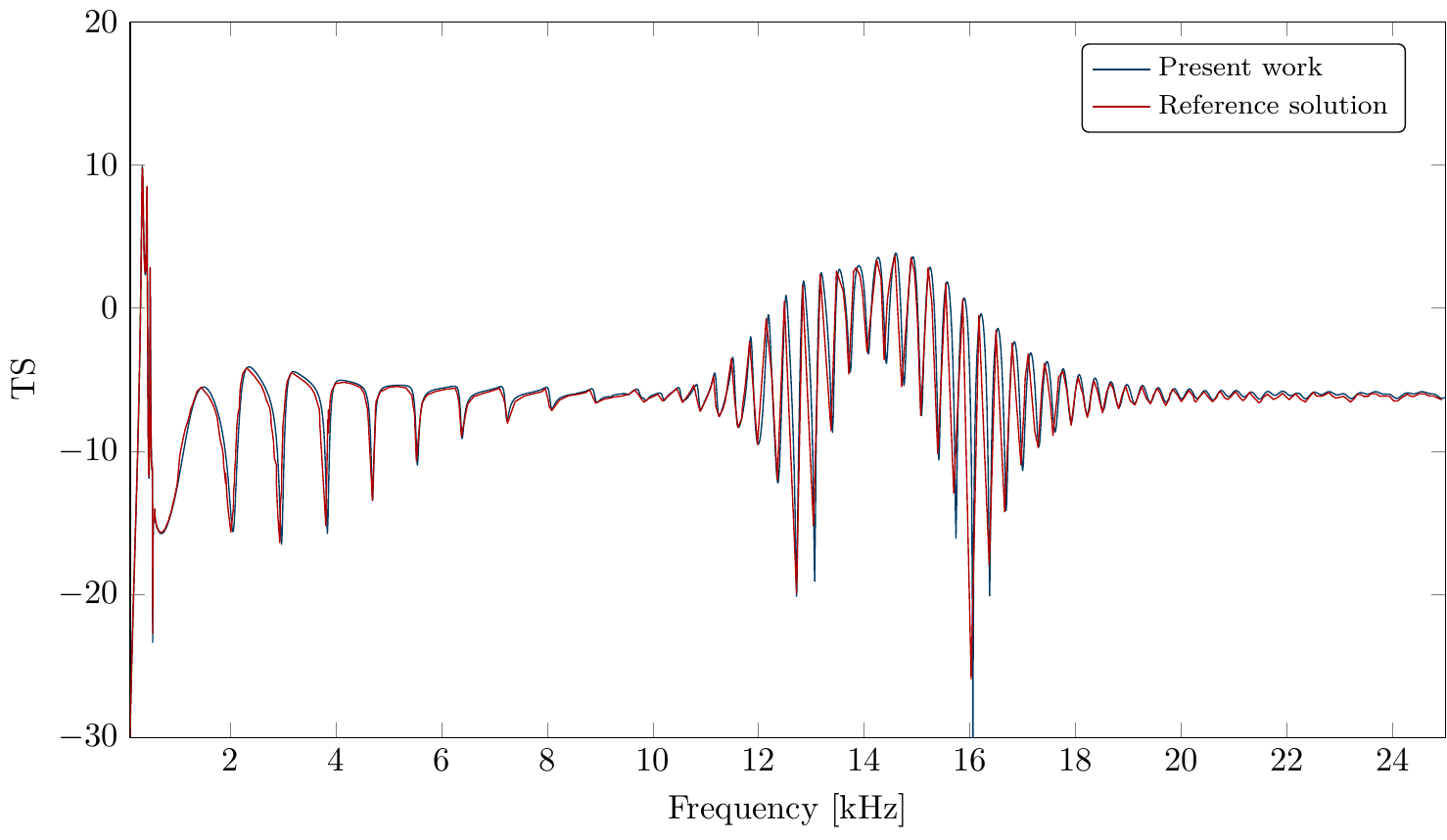}
		\caption{Plot over the frequencies $f\in [0,\SI{25}{kHz}]$.}
		\label{Fig:Skelton2a}
	\end{subfigure}
	\par\bigskip
	\begin{subfigure}[t]{\textwidth}
		\centering
		\includegraphics{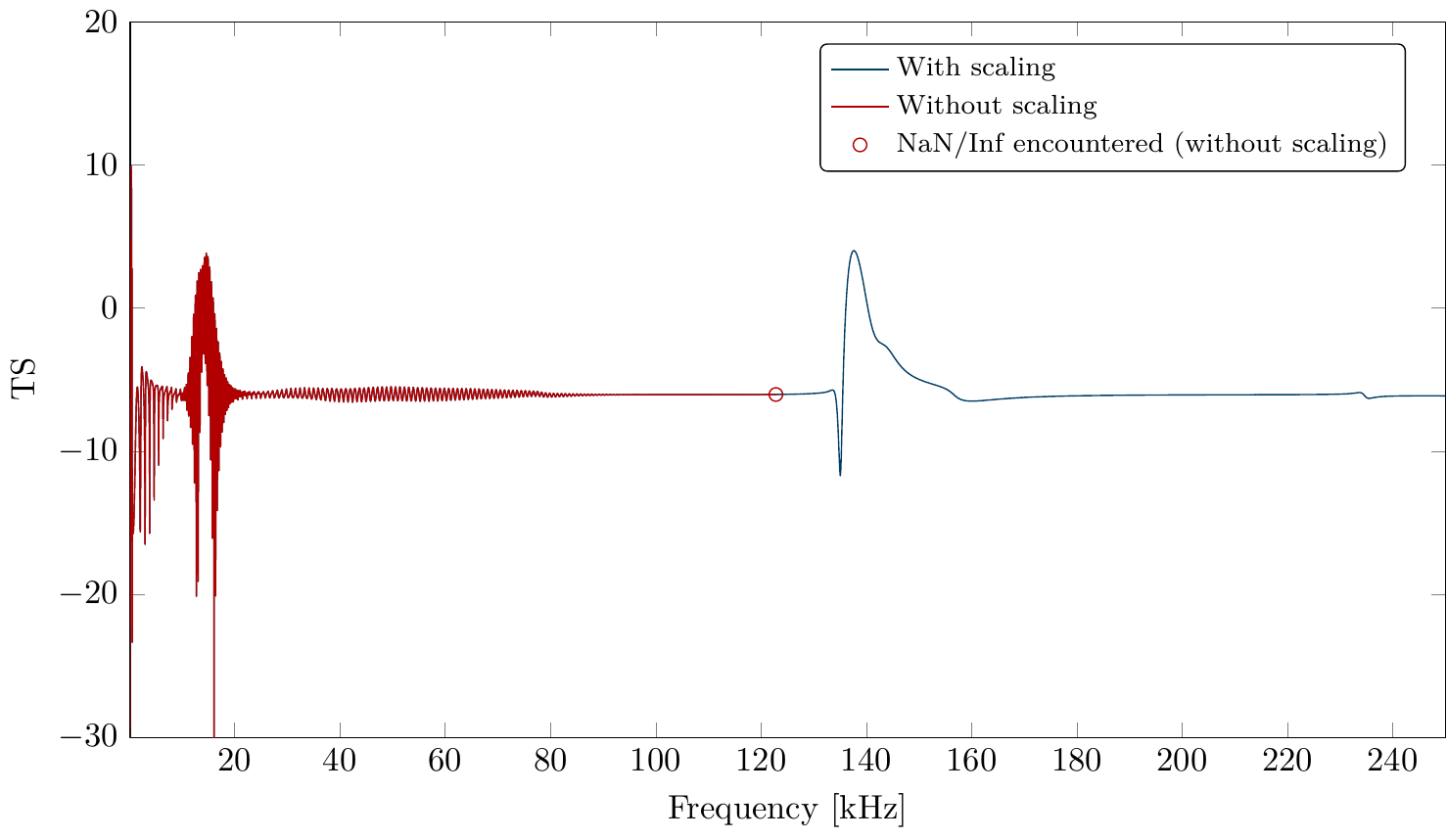}
		\caption{Plot over the frequencies $f\in [0,\SI{250}{kHz}]$.}
		\label{Fig:Skelton2b}
	\end{subfigure}
	\caption{\textbf{Skelton benchmark problem}: A steel shell is implemented with a sound soft boundary condition at its interior and is insonified with a plane wave. The latter plots considers a larger range of frequencies to illustrate the importance of the scaling strategy presented herein.}
\end{figure}

Skelton~\cite{Skelton1997tao} considers a steel spherical shell with a hysteretic loss factor, scattering an incident plane wave directed along the positive $x_3$-axis, and uses the parameters in \Cref{Tab1:Skelton}.
\begin{table}
	\centering
	\caption{\textbf{Skelton parameters:} Parameters for the examples in Figure 10.4 to Figure 10.7 in \cite{Skelton1997tao}.}
	\label{Tab1:Skelton}
	\begin{tabular}{l l}
		\toprule
		Parameter & Description\\
		\midrule
		$R_1 = \SI{1.02}{m}$ & Inner radius of domain 1\\
		$\rho_{1} = \SI{1000}{kg.m^{-3}}$ & Density of water\\
		$c_{1} = \SI{1500}{m.s^{-1}}$ & Speed of sound in water\\
		$R_2 = \SI{1.0}{m}$ & Inner radius of domain 2\\
		$\rho_{2} = \SI{800}{kg.m^{-3}}$ & Density of decoupling\\
		$\tilde{\eta}_{1,2}=\tilde{\eta}_{2,2}= 0.1$ & Loss factor in coating layer\\
		$E_2 = \SI{0.260e7}{Pa}$ & Young's modulus in decoupling\\
		$\nu_2 = 0.460$ & Poisson's ratio in decoupling\\
		$R_3 = \SI{0.98}{m}$ & Inner radius of domain 3\\
		$\rho_{3} = \SI{7700}{kg.m^{-3}}$ & Density of steel\\
		$\tilde{\eta}_{1,3}=\tilde{\eta}_{2,3}= 0.01$ & Loss factor in steel shell\\
		$E_3 = \SI{0.195e12}{Pa}$ & Young's modulus in steel shell\\
		$\nu_3 = 0.290$ & Poisson's ratio in steel shell\\
		\bottomrule
	\end{tabular}
\end{table}
In \Cref{Fig:Skelton} several cases are compared with the data from Figure 10.4 to Figure 10.7 in~\cite{Skelton1997tao}\footnote{The discrepancies probably comes from the fact that the data set is collected by the software \href{https://automeris.io/WebPlotDigitizer/}{WebPlotDigitizer} where a digital scan of Figure 10.4 to Figure 10.7 in~\cite{Skelton1997tao} has been made.}. Good agreements are observed in all cases considered.
The case of an empty steel shell without coating is considered for higher frequencies in~\Cref{Fig:Skelton2a} and higher still in~\Cref{Fig:Skelton2b}. The latter plots illustrate the importance of the scaling strategy presented herein as the solution without scaling (computed from code based on~\cite{Jenserud1990ars}) does not converge to machine epsilon precision due to overflow computations. In~\cite{Skelton1997tao} the equally spaced dips in the range $f \in [\SI{2}{kHz},\SI{10}{kHz}]$ and the ``hump'' round $f=\SI{14}{kHz}$ is explained to stem from first symmetric Lamb wave and first antisymmetric Lamb wave, respectively. However, both of these are actually symmetric Lamb waves, and the phenomena round $f=\SI{140}{kHz}$ (not covered in~\cite{Skelton1997tao}) represents antisymmetric Lamb waves.

\subsection{Hetmaniuk benchmark problem} 
\begin{figure}
	\centering
	\begin{subfigure}[t]{\textwidth}
		\centering
		\includegraphics{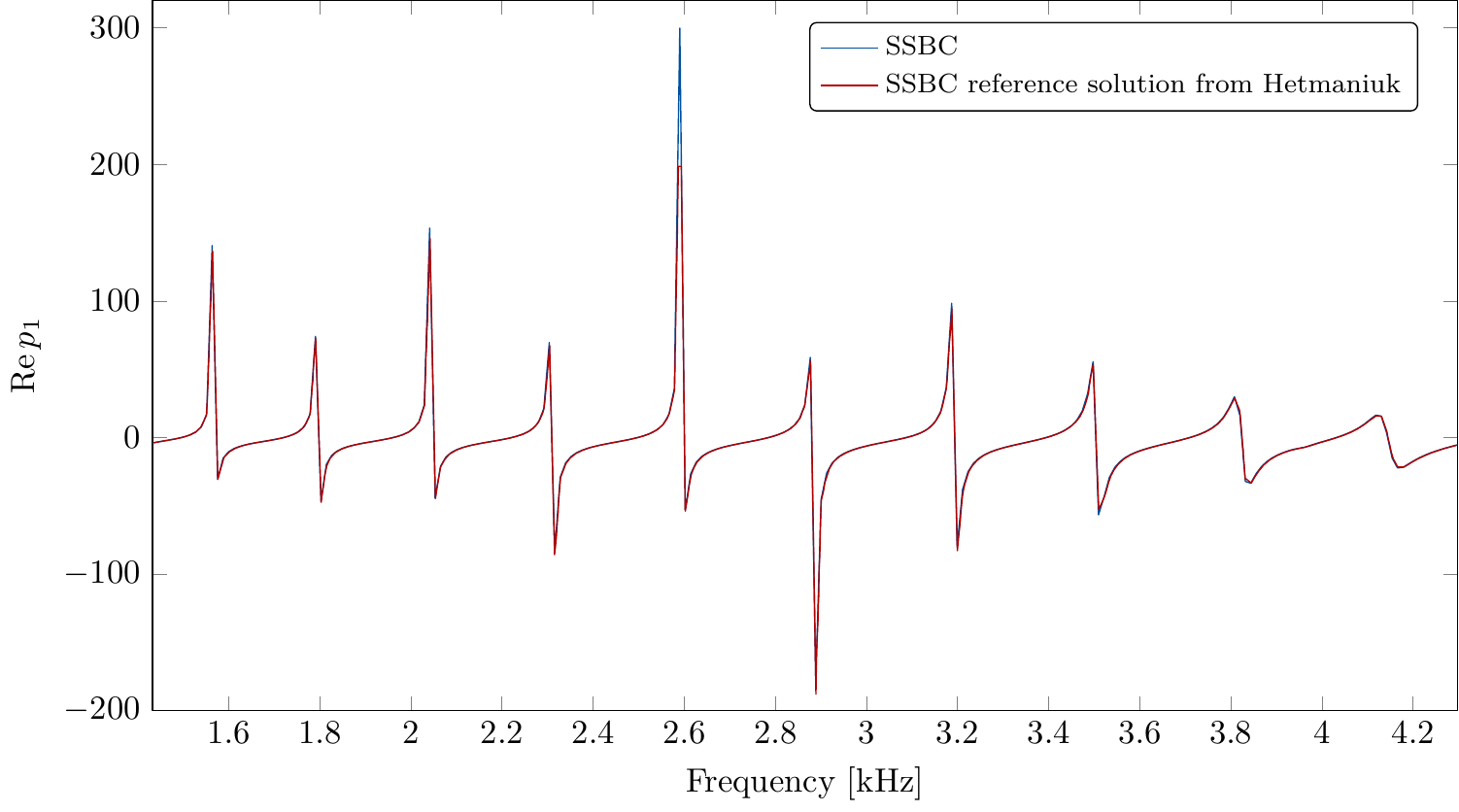}
		\caption{Steel shell excited by a point load.}
		\label{Fig:Hetmaniuk2a}
	\end{subfigure}
	\par\bigskip
	\begin{subfigure}[t]{\textwidth}
		\centering
		\includegraphics{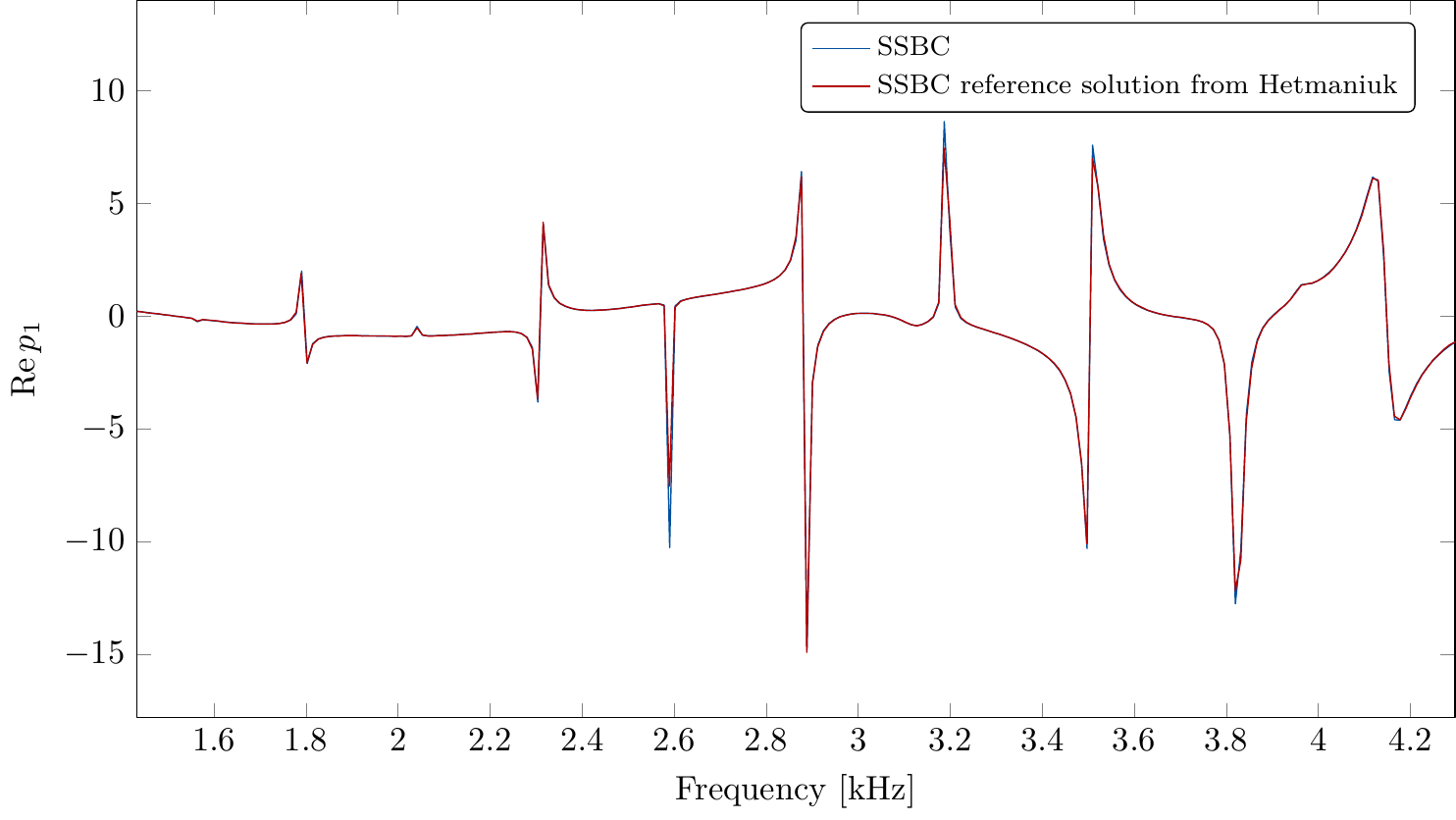}
		\caption{Steel shell excited by a plane wave.}
		\label{Fig:Hetmaniuk2b}
	\end{subfigure}
	\caption{\textbf{Hetmaniuk benchmark problem}: Comparison between present work and a finite element solution from~\cite{Hetmaniuk2012raa}.}
\end{figure}

Hetmaniuk~\cite{Hetmaniuk2012raa} considers a steel spherical shell with the parameters in \Cref{Tab1:Hetmaniuk}.
\begin{table}
	\centering
	\caption{\textbf{Hetmaniuk parameters:} Parameters for the examples in Figure 12 and Figure 17 in \cite{Hetmaniuk2012raa}.}
	\label{Tab1:Hetmaniuk}
	\begin{tabular}{l l}
		\toprule
		Parameter & Description\\
		\midrule
		$R_1 = \SI{1}{m}$ & Inner radius of domain 1\\
		$\rho_{1} = \SI{1000}{kg.m^{-3}}$ & Density of water\\
		$c_{1} = \SI{1500}{m.s^{-1}}$ & Speed of sound in water\\
		$R_2 = \SI{0.95}{m}$ & Inner radius of domain 2\\
		$\rho_{2} = \SI{7850}{kg.m^{-3}}$ & Density of steel\\
		$E_2 = \SI{2.0e12}{Pa}$ & Young's modulus in steel shell\\
		$\nu_2 = 0.3$ & Poisson's ratio in steel shell\\
		\bottomrule
	\end{tabular}
\end{table}
Two load examples are considered: In~\Cref{Fig:Hetmaniuk2a} a point excitation (\Cref{Eq:P_inc_mech}) is considered at $R_2 \vec{d}_{\mathrm{s}}$, and in~\Cref{Fig:Hetmaniuk2b} a plane wave with direction $\vec{d}_{\mathrm{s}}$ is considered. Both plots are compared with the data from Figure 12 and Figure 17 in~\cite{Hetmaniuk2012raa}\footnote{The discrepancies probably comes from the fact that the data set is collected by the software \href{https://automeris.io/WebPlotDigitizer/}{WebPlotDigitizer} where a digital scan of Figure 12 and Figure 17 in~\cite{Hetmaniuk2012raa} has been made.}, and very good agreement is yet again observed.

\subsection{Ayres benchmark problem} 
\begin{figure}
	\centering
	\begin{subfigure}[t]{\textwidth}
		\centering
		\includegraphics{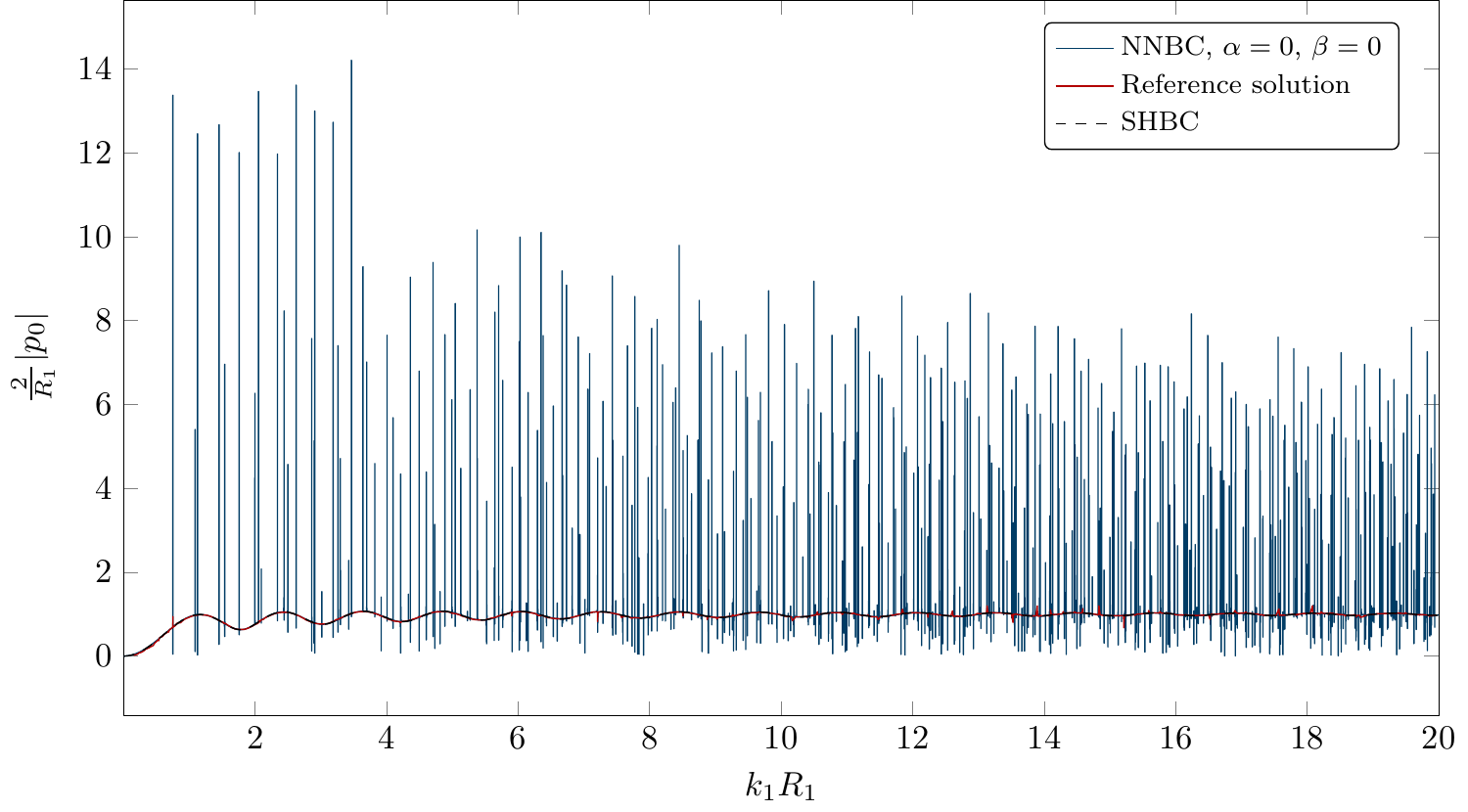}
		\caption{Rubber sphere excited by a plane wave (no damping).}
		\label{Fig:Ayres2a}
	\end{subfigure}
	\par\bigskip
	\begin{subfigure}[t]{\textwidth}
		\centering
		\includegraphics{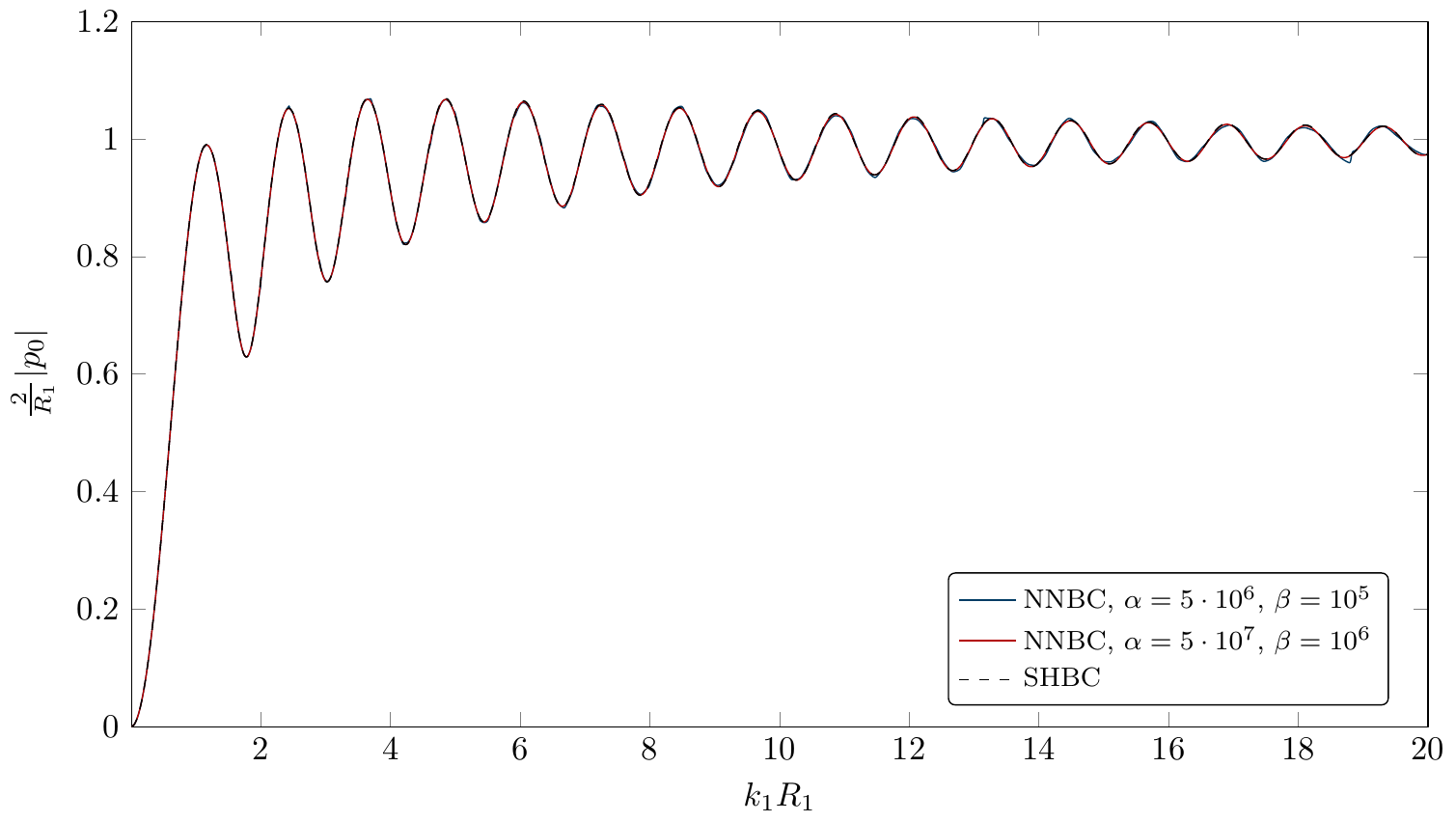}
		\caption{Rubber sphere excited by a plane wave (with damping).}
		\label{Fig:Ayres2b}
	\end{subfigure}
	\caption{\textbf{Ayres benchmark problem}: Comparison between present work and a reference solution from~\cite{Ayres1987ars} for different pairs of $\alpha$ and $\beta$.}
\end{figure}

Ayres~\cite{Ayres1987ars} considers a rubber spherical shell with the parameters in \Cref{Tab1:Ayres}.
\begin{table}
	\centering
	\caption{\textbf{Ayres parameters:} Parameters for the examples in Figure 12 and Figure 17 in \cite{Ayres1987ars}.}
	\label{Tab1:Ayres}
	\begin{tabular}{l l}
		\toprule
		Parameter & Description\\
		\midrule
		$\rho_{1} = \SI{1.2}{kg.m^{-3}}$ & Density of air\\
		$c_{1} = \SI{334}{m.s^{-1}}$ & Speed of sound in air\\
		$\rho_{2} = \SI{1130}{kg.m^{-3}}$ & Density of rubber\\
		$c_{\mathrm{s},1,2} = \SI{1400}{m.s^{-1}}$ & Longitudinal wave velocity\\
		$c_{\mathrm{s},2,2} = \SI{94}{m.s^{-1}}$ & Transverse wave velocity\\
		\bottomrule
	\end{tabular}
\end{table}
The loss factors for the rubber sphere is here defined by
\begin{equation}
	\tilde{\eta}_{1,2} = \frac{\alpha + 2\beta}{\rho_{2}c_{\mathrm{s},1,2}^2},\quad\text{and}\quad\tilde{\eta}_{2,2} = \frac{\beta}{\rho_{2}c_{\mathrm{s},2,2}^2}.
\end{equation}
Three pairs of $\alpha$ and $\beta$ was considered and the resulting scaled far-field pattern is plotted in~\Cref{Fig:Ayres2a} and~\Cref{Fig:Ayres2b}. Both plots are compared with the data from Figure 1 in~\cite{Ayres1987ars}\footnote{The discrepancies probably comes from the fact that the data set is collected by the software \href{https://automeris.io/WebPlotDigitizer/}{WebPlotDigitizer} where a digital scan of Figure 1 in~\cite{Ayres1987ars} has been made.} but is sampled a lot more closely to reveal more eigenmodes, and good agreement is yet again observed.

\section{Conclusions}
\label{Sec1:conclusions}
In this work the solution in~\cite{Venas2019e3s} has been generalized and extended with new boundary conditions and loads, and generalized to include materials with damping. Moreover, a scaling strategy is presented which extracts exponential behaviour from the Bessel functions in order to deal with overflow issues. The problem is significantly mitigated and extends the domain of computation by an order of magnitude for the frequency. Using quadruple precision, the problem is eliminated for practical applications at the expense of the computational efficiency. This significantly extends the range for the parameters at which the solution can be computationally evaluated.

Several numerical examples has been investigated and verified against reference solutions in the literature. The example from Skelton illustrates the importance of the scaling strategy presented herein as previously hidden important characteristics has now been reviled.

\section*{Acknowledgments}
This work was supported by SINTEF Digital.
% and by the Norwegian Defence Research Establishment.

\appendix
\clearpage
\section{Derived functions}
\label{Sec1:derivedFunctions}
For completeness we list the resulting solutions for the displacement and the stress fields (cf.~\cite{Venas2019e3s})
\begin{equation}\label{Eq1:u_rgen}
	u_{\mathrm{r},m} = \frac{1}{r}\sum_{n=0}^\infty Q_n^{(0)}(\vartheta)\left[A_n^{(i)}S_{1,n}^{(i)}(\xi_m(r))+B_n^{(i)}T_{1,n}^{(i)}(\eta_m(r))\right]
\end{equation}
and
\begin{equation}\label{Eq1:u_tgen}
	u_{\upvartheta,m} = \frac{1}{r}\sum_{n=0}^\infty Q_n^{(1)}(\vartheta)\left[A_n^{(i)}S_{2,n}^{(i)}(\xi_m(r))+B_n^{(i)}T_{2,n}^{(i)}(\eta_m(r))\right]
\end{equation}
where
\begin{align*}
	S_{1,n}^{(i)}(\xi) &= nZ_n^{(i)}(\xi)-\xi Z_{n+1}^{(i)}(\xi)\\ 
	T_{1,n}^{(i)}(\eta) &= -n(n+1)Z_n^{(i)}(\eta)\\
	S_{2,n}^{(i)}(\xi) &= Z_n^{(i)}(\xi)\\
	T_{2,n}^{(i)}(\eta) &= -(n+1)Z_n^{(i)}(\eta) + \eta Z_{n+1}^{(i)}(\eta).
\end{align*}
The derivatives of the displacement fields are found to be
\begin{align}
	\pderiv{u_{\mathrm{r},m}}{r} &= \frac{1}{r^2}\sum_{n=0}^\infty Q_n^{(0)}(\vartheta)\left[A_n^{(i)}S_{3,n}^{(i)}(\xi_m(r))+B_n^{(i)}T_{3,n}^{(i)}(\eta_m(r))\right]\\
	\pderiv{u_{\upvartheta}}{r} &= \frac{1}{r^2}\sum_{n=0}^\infty Q_n^{(1)}(\vartheta)\left[A_n^{(i)}S_{4,n}^{(i)}(\xi_m(r))+B_n^{(i)}T_{4,n}^{(i)}(\eta_m(r))\right]\\
	\pderiv{u_{\mathrm{r},m}}{\vartheta} &= \frac{1}{r}\sum_{n=0}^\infty Q_n^{(1)}(\vartheta)\left[A_n^{(i)}S_{1,n}^{(i)}(\xi_m(r))+B_n^{(i)}T_{1,n}^{(i)}(\eta_m(r))\right]\\
	\pderiv{u_{\upvartheta,m}}{\vartheta} &= \frac{1}{r}\sum_{n=0}^\infty Q_n^{(2)}(\vartheta)\left[A_n^{(i)}S_{2,n}^{(i)}(\xi_m(r))+B_n^{(i)}T_{2,n}^{(i)}(\eta_m(r))\right]
\end{align}
where
\begin{align*}
	S_{3,n}^{(i)}(\xi) &=  (n^2-\xi^2-n)Z_n^{(i)}(\xi) + 2\xi Z_{n+1}^{(i)}(\xi)\\ 
	T_{3,n}^{(i)}(\eta) &= -n(n+1)\left[(n-1)Z_n^{(i)}(\eta) - \eta Z_{n+1}^{(i)}(\eta)\right]\\
	S_{4,n}^{(i)}(\xi) &= (n-1)Z_n^{(i)}(\xi)-\xi Z_{n+1}^{(i)}(\xi)\\ 
	T_{4,n}^{(i)}(\eta) &= (\eta^2-n^2+1)Z_n^{(i)}(\eta) -\eta Z_{n+1}^{(i)}(\eta).
\end{align*}
The stress fields are given by (cf.~\cite{Venas2019e3s})\footnote{Note that formulas in~\cite[Eq. (B.8)]{Venas2019e3s} are related to~\cite[Eqs. (23a) and (23b)]{Hasheminejad2005asf} by
\begin{align*}
    \sigma_{\mathrm{rr},m} &= \left(K_m+\frac{4G_m}{3}\right)\varepsilon_{\mathrm{rr},m} + \left(K_m-\frac{2G_m}{3}\right)\varepsilon_{\upvartheta\upvartheta,m} + \left(K_m-\frac{2G_m}{3}\right)\varepsilon_{\upvarphi\upvarphi,m}\\
    &= -p_m + \left(\mu_{\mathrm{b},m}-\frac23\mu_m\right)\nabla\cdot\vec{v}_m + 2\mu_m\pderiv{v_{\mathrm{r},m}}{r}\\
    \sigma_{\mathrm{r}\upvartheta,m} &= G_m\left(\frac{1}{r}\pderiv{u_{\mathrm{r},m}}{\vartheta} + \pderiv{u_{\vartheta,m}}{r} - \frac{u_\vartheta}{r}\right)\\
     &= \mu_m\left(\frac{1}{r}\pderiv{v_{\mathrm{r},m}}{\vartheta} + \pderiv{v_{\vartheta,m}}{r} - \frac{v_\vartheta}{r}\right)\\
\end{align*}
where
\begin{align*}
    \varepsilon_{\mathrm{rr},m} = \pderiv{u_{\mathrm{r},m}}{r},\quad\varepsilon_{\upvartheta\upvartheta,m} = \frac{1}{r}\left(\pderiv{u_{\upvartheta,m}}{\upvartheta} + u_{\mathrm{r},m}\right),\quad \varepsilon_{\upvarphi\upvarphi,m} = \frac{1}{r\sin\theta}\left(\pderiv{u_{\upvarphi,m}}{\upvarphi} + u_{\mathrm{r},m}\sin\vartheta + u_{\upvartheta,m}\cos\vartheta\right).
\end{align*}
}
\begin{align}\label{Eq1:stressFieldComponents1}
	\sigma_{\mathrm{r}\mathrm{r},m} &= \frac{2G_m}{r^2}\sum_{n=0}^\infty Q_n^{(0)}(\vartheta)\left[A_n^{(i)} S_{5,n}^{(i)}(\xi_m(r); a_m, b_m) + B_n^{(i)} T_{5,n}^{(i)}(\eta_m(r))\right]\\\label{Eq1:stressFieldComponents2}
	\sigma_{\upvartheta\upvartheta,m} &= \frac{2G_m}{r^2}\sum_{n=0}^\infty\left\{Q_n^{(0)}(\vartheta)\left[A_n^{(i)} S_{6,n}^{(i)}(\xi_m(r); a_m, b_m) + B_n^{(i)} T_{6,n}^{(i)}(\eta_m(r))\right] \right.\\
	&\qquad\qquad\qquad\left.+  Q_n^{(2)}(\vartheta)\left[A_n^{(i)} S_{2,n}^{(i)}(\xi_m(r)) + B_n^{(i)} T_{2,n}^{(i)}(\eta_m(r))\right]\right\}\\\label{Eq1:stressFieldComponents3}
	\sigma_{\upvarphi\upvarphi,m} &= \frac{2G_m}{r^2}\sum_{n=0}^\infty\left\{Q_n^{(0)}(\vartheta)\left[A_n^{(i)} S_{6,n}^{(i)}(\xi_m(r); a_m, b_m) + B_n^{(i)} T_{6,n}^{(i)}(\eta_m(r))\right] \right.\\
	&\qquad\qquad\qquad\left.+  Q_n^{(1)}(\vartheta)\cot(\vartheta)\left[A_n^{(i)} S_{2,n}^{(i)}(\xi_m(r)) + B_n^{(i)} T_{2,n}^{(i)}(\eta_m(r))\right]\right\}\\\label{Eq1:stressFieldComponents4}
	\sigma_{\upvartheta\upvarphi,m} &= 0\\\label{Eq1:stressFieldComponents5}
	\sigma_{\mathrm{r}\upvarphi,m} &= 0\\\label{Eq1:stressFieldComponents6}
	\sigma_{\mathrm{r}\upvartheta,m} &= \frac{2G_m}{r^2}\sum_{n=0}^\infty Q_n^{(1)}(\vartheta)\left[A_n^{(i)} S_{7,n}^{(i)}(\xi_m(r)) + B_n^{(i)} T_{7,n}^{(i)}(\eta_m(r))\right]
\end{align}
where
\begin{align}\label{Eq1:stress}
\begin{split}
	S_{5,n}^{(i)}(\xi; a, b) &= \left[n^2-n-\frac{1}{2}\left(\frac{b}{a}\right)^2\xi^2\right] Z_n^{(i)}(\xi) + 2\xi Z_{n+1}^{(i)}(\xi)\\
	T_{5,n}^{(i)}(\eta) &= -n(n+1)\left[(n-1)Z_n^{(i)}(\eta) - \eta Z_{n+1}^{(i)}(\eta)\right]\\
	S_{6,n}^{(i)}(\xi; a, b) &= \left[n-\frac{1}{2}\left(\frac{b}{a}\right)^2\xi^2+\xi^2\right] Z_n^{(i)}(\xi) - \xi Z_{n+1}^{(i)}(\xi)\\
	T_{6,n}^{(i)}(\eta) &= -n(n+1)Z_n^{(i)}(\eta)\\
	S_{7,n}^{(i)}(\xi) &= (n-1)Z_n^{(i)}(\xi) -\xi Z_{n+1}^{(i)}(\xi)\\
	T_{7,n}^{(i)}(\eta) &= -\left(n^2-1-\frac{1}{2}\eta^2\right)Z_n^{(i)}(\eta) - \eta Z_{n+1}^{(i)}(\eta).
	\end{split}
\end{align}
\section{The incident wave}
\label{Sec1:incidentWave}
At the interfaces it is assumed that the incident wave may be decomposed as
\begin{equation*}
    \phi_{\mathrm{inc},m} = \sum_{n=0}^\infty Q_n^{(0)}(\vartheta) A_{\mathrm{inc},m,n}^{(i)}Z_n^{(i)}(\xi_m(r)),
\end{equation*}
where the coefficients are found from
\begin{equation*}
    A_{\mathrm{inc},m,n}^{(i)}Z_n^{(i)}(\xi_m(r)) = \frac{2n+1}{2}\int_0^\PI \phi_{\mathrm{inc},m}(r,\vartheta) Q_n^{(0)}(\vartheta)\sin\vartheta\idiff\vartheta.
\end{equation*}
A plane wave has such an expansion (it is most sensible to imposed this only in the outermost unbounded domain, that is $\phi_{\mathrm{inc},m}=0$ for $m>1$)
\begin{equation*}
    \phi_{\mathrm{inc},1} = \Phi_{\mathrm{inc},1}\euler^{\imag a_1 r\cos\vartheta} = \Phi_{\mathrm{inc},1}\sum_{n=0}^\infty Q_n^{(0)}(\vartheta) (2n+1)\imag^n Z_n^{(1)}(\xi_1(r)),
\end{equation*}
such that $A_{\mathrm{inc},1,n}^{(1)}=\Phi_{\mathrm{inc},1}(2n+1)\imag^n$ and $A_{\mathrm{inc},1,n}^{(i)}=0$ for $i=2,3$.

The incident wave due to a point source located at $\vec{x}_{\mathrm{s}} = r_{\mathrm{s}}\vec{e}_3$ (in domain $m$ such that $R_m < r_{\mathrm{s}} < R_{m-1}$) can be more efficiently evaluated then what was presented in~\cite{Venas2019e3s}. Consider the incident wave\footnote{The relation between $\Phi_{\mathrm{inc},m}$ and the amplitude of the incident wave in a non-viscous fluid used in~\cite{Venas2019e3s}, $P_{\mathrm{inc},m}$, is given by $P_{\mathrm{inc},m} = \rho_m\omega^2\Phi_{\mathrm{inc},m}$}
\begin{equation}
	\phi_{\mathrm{inc},m} = \Phi_{\mathrm{inc},m}\frac{\euler^{\imag a_m |\vec{x}-\vec{x}_{\mathrm{s}}|}}{|\vec{x}-\vec{x}_{\mathrm{s}}|},\quad |\vec{x}-\vec{x}_{\mathrm{s}}| = \sqrt{r^2-2r_{\mathrm{s}}r\cos\vartheta + r_{\mathrm{s}}^2}.
\end{equation}
Using~\cite[10.1.45 and 10.1.46]{Abramowitz1965hom}
\begin{align}
	\frac{\sin a_m R}{a_m R} &= \sum_{n=0}^\infty (2n+1)\besselj_n(a_m r)\besselj_n(a_m r_{\mathrm{s}}) \legendre_n(\cos\vartheta)\\
	-\frac{\cos a_m R}{a_m R} &= \sum_{n=0}^\infty (2n+1)\besselj_n(a_m r)\bessely_n(a_m r_{\mathrm{s}}) \legendre_n(\cos\vartheta),\quad |r| < |r_{\mathrm{s}}|\label{Eq:coskRkR}
\end{align}
where
\begin{equation*}
	R = \sqrt{r^2-2r r_{\mathrm{s}}\cos\theta+r_{\mathrm{s}}^2}
\end{equation*}
we get
\begin{equation}
	\phi_{\mathrm{inc},m} = \Phi_{\mathrm{inc},m}\imag a_m \sum_{n=0}^\infty (2n+1) \legendre_n(\cos\vartheta)\begin{cases}
		\besselj_n(a_m r)\hankel_n^{(1)}(a_m r_{\mathrm{s}}) & r < r_{\mathrm{s}}\\
		\besselj_n(a_m r_{\mathrm{s}})\hankel_n^{(1)}(a_m r) & r > r_{\mathrm{s}}.\\
	\end{cases}
\end{equation}
% Hence, the coefficients
% \begin{align*}
% 	F_{n,\tilde{m}}^{(1)} &= \Phi_{\mathrm{inc},m}\imag a_m (2n+1)  \begin{cases}
% 				\besselj_n(k_1 R_{\tilde{m}}m)\hankel_n^{(1)}(a_m r_{\mathrm{s}}) & R_{\tilde{m}} < r_{\mathrm{s}}\\
% 				\besselj_n(k_1 r_{\mathrm{s}})\hankel_n^{(1)}(a_m R_{\tilde{m}}) & R_{\tilde{m}} > r_{\mathrm{s}}
% 				\end{cases}\\
% 	F_{n,\tilde{m}}^{(2)} &= \Phi_{\mathrm{inc},m}\imag a_m^2 (2n+1) \begin{cases}
% 				\besselj_n'(a_m R_{\tilde{m}})\hankel_n^{(1)}(a_m r_{\mathrm{s}}) & R_{\tilde{m}} < r_{\mathrm{s}}\\
% 				\besselj_n(a_m r_{\mathrm{s}})\left(\hankel_n^{(1)}\right)'(a_m R_{\tilde{m}}) & R_{\tilde{m}} > r_{\mathrm{s}}
% 				\end{cases}
% \end{align*}
% for the conditions at the interface $\tilde{m}\in\{m-1,m\}$.

It is also possible to implement a point force mechanical excitation located at $\vec{x}_{\mathrm{s}} = r_{\mathrm{s}}\vec{e}_3$ where $r_{\mathrm{s}}=R_m$ for a given interface $m$~\cite{Skelton1997tao}
\begin{equation}\label{Eq:P_inc_mech}
	\phi_{\mathrm{inc},m}\vert_{r=R_m} = \Phi_{\mathrm{inc},m}\frac{\delta(\vartheta)}{2\PI R_m^2\sin\vartheta} = \Phi_{\mathrm{inc},m}\sum_{n=0}^\infty Q_n^{(0)}(\vartheta) \frac{2n+1}{4\PI R_m^2}
\end{equation}
where $\delta$ is the Dirac delta function. 

A surface excitation over the axisymmetric region defined by the limits $\vartheta\in[\vartheta_{\mathrm{s},1},\vartheta_{\mathrm{s},2}]$, $\varphi\in[0,2\PI]$ at $r=R_m$ for a given interface $m$ can also be implemented
\begin{align*}
	\phi_{\mathrm{inc},m}\vert_{r=R_m} &= \Phi_{\mathrm{inc},m}[\heaviside(\vartheta-\vartheta_{\mathrm{s},1}) - \heaviside(\vartheta-\vartheta_{\mathrm{s},2})] \\
	&= \frac{\Phi_{\mathrm{inc},m}}{2}\sum_{n=0}^\infty Q_n^{(0)}(\vartheta) \left[\legendre_{n-1}(\cos\vartheta_{\mathrm{s},2})-\legendre_{n+1}(\cos\vartheta_{\mathrm{s},2}) - (\legendre_{n-1}(\cos\vartheta_{\mathrm{s},1})-\legendre_{n+1}(\cos\vartheta_{\mathrm{s},1}))\right]
\end{align*}
where $\heaviside$ is the Heaviside function.

%\section*{References}
\bibliographystyle{TK_CM} % elsarticle-num2 is the same as elsarticle-num, but without the display of url and doi (the title contains the url info.)

\bibliography{references_zotero}

\end{document}